\documentclass[zpreprint,zbstepj]{./LaTeX/zeus/zeus_paper}
\usepackage[english]{babel}

\newcommand{\ZcoosysB}{%
The ZEUS coordinate system is a right-handed Cartesian system, with the $Z$
axis pointing in the proton beam direction, referred to as the ``forward
direction'', and the $X$ axis pointing left towards the centre of HERA.
The coordinate origin is at the nominal interaction point.\xspace}
\newcommand{\Zpsrap}{%
The pseudorapidity is defined as $\eta=-\ln\left(\tan\frac{\theta}{2}\right)$,
where the polar angle, $\theta$, is measured with respect to the proton beam
direction.\xspace}

\newcommand{\ZcoosysfnBeta}{\footnote{\ZcoosysB\Zpsrap}}
\newcommand{\Zdetdesc}{%
A detailed description of the ZEUS detector can be found 
elsewhere~\cite{zeus:1993:bluebook}. A brief outline of the 
components that are most relevant for this analysis is given
below.\xspace}
\newcommand{\Zctddesc}[1]{%
Charged particles are tracked in the central tracking detector (CTD)~\citeCTD,
which operates in a magnetic field of $1.43\Tesla$ provided by a thin 
superconducting solenoid. The CTD consists of 72~cylindrical drift chamber 
layers, organised in nine superlayers covering the polar-angle#1 region 
\mbox{$15^\circ<\theta<164^\circ$}. The transverse-momentum resolution for
full-length tracks is $\sigma(p_T)/p_T=0.0058p_T\oplus0.0065\oplus0.0014/p_T$,
with $p_T$ in $\Gev$.}
\newcommand{\Zcaldesc}{%
The high-resolution uranium--scintillator calorimeter (CAL)~\citeCAL consists 
of three parts: the forward (FCAL), the barrel (BCAL) and the rear (RCAL)
calorimeters. Each part is subdivided transversely into towers and
longitudinally into one electromagnetic section (EMC) and either one (in RCAL)
or two (in BCAL and FCAL) hadronic sections (HAC). The smallest subdivision of
the calorimeter is called a cell.  The CAL energy resolutions, as measured under
test-beam conditions, are $\sigma(E)/E=0.18/\sqrt{E}$ for electrons and
$\sigma(E)/E=0.35/\sqrt{E}$ for hadrons, with $E$ in $\Gev$.
The timing resolution of the CAL is better than 1$\,$ns for energy
deposits greater than $4.5\,$GeV.}

\chardef\usc=95
\chardef\til=126
\catcode`\@=11 
\DeclareRobustCommand\xdotspace{\futurelet\@let@token\@xdotspace}
\def\@xdotspace{%
  \ifx\@let@token.\else
  \ifx\@let@token\bgroup.\else
  \ifx\@let@token\egroup.\else
  \ifx\@let@token\/.\else
  \ifx\@let@token\ .\else
  \ifx\@let@token~.\else
  \ifx\@let@token!.\else
  \ifx\@let@token,.\else
  \ifx\@let@token:.\else
  \ifx\@let@token;.\else
  \ifx\@let@token?.\else
  \ifx\@let@token/.\else
  \ifx\@let@token'.\else
  \ifx\@let@token).\else
  \ifx\@let@token-.\else
  \ifx\@let@token\@xobeysp.\else
  \ifx\@let@token\space.\else
  \ifx\@let@token\@sptoken.\else
   .\space
   \fi\fi\fi\fi\fi\fi\fi\fi\fi\fi\fi\fi\fi\fi\fi\fi\fi\fi}
\catcode`\@=12 

\newcommand{\stru}[2]{%
   \relax\ifmmode\hbox{\vrule height#1 depth#2 width0pt}%
   \else\vrule height#1 depth#2 width0pt\fi}

\newcommand{\Ronum}[1]{\uppercase\expandafter{\romannumeral#1}}
\newcommand{\ronum}[1]{\expandafter{\romannumeral#1}}
\DeclareRobustCommand{\LaTeXZ}{%
  \LaTeX\kern-.05em4\kern-.1em
  {\raisebox{-0.2ex}{$\scriptstyle\text{ZEUS}$}}\xspace}

\newcommand{\Sect}[1]{Section~\ref{sec-#1}}

\DeclareMathAlphabet{\mathbf}{OT1}{cmr}{bx}{sl}
\newcommand{\eVdist}{\kern-0.06667em}

\newcommand{\Gev}{{\text{Ge}\eVdist\text{V\/}}}

\newcommand{\mev}{{\,\text{Me}\eVdist\text{V\/}}}
\newcommand{\gev}{{\,\text{Ge}\eVdist\text{V\/}}}

\newcommand{\met}{\,\text{m}}

\newcommand{\Tesla}{\,\text{T}}

\newcommand{\slashfrac}[2]{%
  \raisebox{0.5ex}{\ensuremath #1}\kern-0.12em/\kern-0.08em
  \raisebox{-.8ex}{\ensuremath #2}}

\newcommand{\sqr}[3]{%
    {\vcenter{\hrule height.#3ex\hbox{\vrule width.#2ex height#1ex
     \kern#1ex\vrule width.#3ex}\hrule height.#2ex}}}

\catcode`\@=11 
\newcommand{\parenbar}{\mathpalette\p@renb@r}
\def\p@renb@r#1#2{\vbox{%
  \ifx#1\scriptscriptstyle \dimen@.7em\dimen@ii.2em\else
  \ifx#1\scriptstyle \dimen@.8em\dimen@ii.25em\else
  \dimen@1em\dimen@ii.4em\fi\fi \offinterlineskip
  \ialign{\hfill##\hfill\cr
    \vbox{\hrule width\dimen@ii}\cr
    \noalign{\vskip-.3ex}%
    \hbox to\dimen@{$\mathchar300\hfil\mathchar301$}\cr
    \noalign{\vskip-.3ex}%
    $#1#2$\cr}}}
\catcode`\@=12 

\newcommand{\diff}{{\rm d}}

\newcommand{\IP}{{\rm I$\kern-0.01667em$P}\xspace}

\newcommand{\smax}{{\rm max}}

\mathchardef\qsm=63
\mathchardef\pls=43
\mathchardef\mns=512
\mathchardef\plm=518
\mathchardef\eql=61
\mathchardef\smallleft=300
\mathchardef\smallright=301
\mathchardef\les=316
\mathchardef\gre=318
\mathchardef\leq=532
\mathchardef\grq=533

\catcode`\@=11 
\newcounter{pict@width}
\newcounter{pict@height}
\newlength{\pict@scale}
\newcommand{\psfigadd}[4]{%
\setcounter{pict@width}{1*\ratio{#2+\pict@scale/2}{\pict@scale}}
\setcounter{pict@height}{1*\ratio{#3+\pict@scale/2}{\pict@scale}}
\setlength{\unitlength}{\pict@scale}
\hbox to #2{\hspace{-\fill}\begin{picture}(\thepict@width,\thepict@height)
\put(0,0){\psfig{figure=#1,width=#2,height=#3,clip=}}
\SetScale{0.283466457}
\SetWidth{1.763889}
{#4}
\end{picture}}
}
\newcounter{pict@widthfst}
\newcounter{pict@widthscd}
\newcounter{pict@widthtot}
\newcommand{\psfigaddtwo}[7]{%
\setcounter{pict@widthfst}{1*\ratio{#2+\pict@scale/2}{\pict@scale}}
\setcounter{pict@widthscd}{1*\ratio{#2+#4+\pict@scale/2}{\pict@scale}}
\setcounter{pict@widthtot}{1*\ratio{#2+#4+#6+\pict@scale/2}{\pict@scale}}
\setcounter{pict@height}{1*\ratio{#3+\pict@scale/2}{\pict@scale}}
\setlength{\unitlength}{\pict@scale}
\hbox{\hspace{-\fill}\begin{picture}(\thepict@widthtot,\thepict@height)
\put(0,0){\psfig{figure=#1,width=#2,height=#3,clip=}}
\put(\thepict@widthscd,0){\psfig{figure=#5,width=#6,height=#3,clip=}}
\SetScale{0.283466457}
\SetWidth{1.763889}
{#7}
\end{picture}}
}
\newcommand{\psfigror}[4]{%
\setcounter{pict@width}{1*\ratio{#2+\pict@scale/2}{\pict@scale}}
\setcounter{pict@height}{1*\ratio{#3+\pict@scale/2}{\pict@scale}}
\setlength{\unitlength}{\pict@scale}
\hbox{\begin{picture}(\thepict@width,\thepict@height)
\put(0,\thepict@height){\psfig{figure=#1,width=#3,height=#2,clip=,angle=270}}
\SetScale{0.283466457}
\SetWidth{1.763889}
{#4}
\end{picture}}
}
\newcommand{\psfigrol}[4]{%
\setcounter{pict@width}{1*\ratio{#2+\pict@scale/2}{\pict@scale}}
\setcounter{pict@height}{1*\ratio{#3+\pict@scale/2}{\pict@scale}}
\setlength{\unitlength}{\pict@scale}
\hbox{\begin{picture}(\thepict@width,\thepict@height)
\put(0,0){\psfig{figure=#1,width=#3,height=#2,clip=,angle=90}}
\SetScale{0.283466457}
\SetWidth{1.763889}
{#4}
\end{picture}}
}
\catcode`\@=12 
\newlength\listtextwidth

\catcode`\@=11 
\newlength{\@tabfninsert}
\newlength{\@tabfnwidth}
\newcommand{\tabfootnote}[2]{%
  \setlength{\@tabfninsert}{0.8em}
  \setlength{\@tabfnwidth}{\textwidth}
  \addtolength{\@tabfnwidth}{-\@tabfninsert}
  \addtolength{\@tabfnwidth}{-0.4em}
  \noindent\makebox[\@tabfninsert][r]{\footnotesize$^{#1}$\hfil}\hfill%
  \parbox[t]{\@tabfnwidth}{\footnotesize #2\hfill}}
\catcode`\@=12 

\newcommand {\dstar} {D^{\ast \pm}}
\newcommand {\etam} {\eta_\smax}

\newcommand {\pom} {I\!\!P}
\newcommand {\pomsub} {{\scriptscriptstyle \pom}}

\newcommand {\xpom} {x_{\pomsub}}

\newcommand{\ccb}        {\mbox{$c\bar{c}$}}

\newcommand{\dspm}       {\mbox{$D^{\ast \pm}$}}
\newcommand{\dz}         {\mbox{$D^{0}$}}

\newcommand{\fcds}       {\mbox{$f(c \rightarrow D^{\ast +})$}}
\def\et10{ E_{\perp}^{\theta > 10^\circ}}
\def\citeCTD{{\cite{%
nim:a279:290,*npps:b32:181,*nim:a338:254%
}}\xspace}
\def\citeCAL{{\cite{%
nim:a309:77,*nim:a309:101,*nim:a321:356,*nim:a336:23%
}}\xspace}

\begin{document}
\prepnum{{DESY--03--094}}

\title{
  Measurement of the
  open-charm contribution to the diffractive proton structure function}                                                       
                    
\author{ZEUS Collaboration}
\date{July 2003}

\abstract{
Production of $D^{*\pm}(2010)$ mesons in diffractive deep inelastic 
scattering has been measured with the ZEUS detector at HERA using an 
integrated luminosity of 82~pb$^{-1}$. Diffractive events were
identified by the presence of a large rapidity gap in the final state. 
Differential cross sections have been measured in the kinematic region 
$1.5 < Q^2 < 200\,$GeV$^2$, $0.02 < y < 0.7$, $\mbox{$x_{I\hspace{-0.2em}P}$}~<~0.035$,
$\beta < 0.8$, 
$p_T(D^{*\pm}) > 1.5\,$GeV and $|\eta(D^{*\pm})| < 1.5$. The measured
cross sections are compared to theoretical predictions.
The results are presented in terms of the
open-charm contribution to the diffractive proton structure function.
The data demonstrate a strong sensitivity to the diffractive parton
densities.
}

\makezeustitle

\def\3{\ss}                                                                                        
\pagenumbering{Roman}                                                                              
                                                   %
\begin{center}                                                                                     
{                      \Large  The ZEUS Collaboration              }                               
\end{center}                                                                                       
  S.~Chekanov,                                                                                     
  M.~Derrick,                                                                                      
  D.~Krakauer,                                                                                     
  J.H.~Loizides$^{   1}$,                                                                          
  S.~Magill,                                                                                       
  B.~Musgrave,                                                                                     
  J.~Repond,                                                                                       
  R.~Yoshida\\                                                                                     
 {\it Argonne National Laboratory, Argonne, Illinois 60439-4815}~$^{n}$                            
\par \filbreak                                                                                     
  M.C.K.~Mattingly \\                                                                              
 {\it Andrews University, Berrien Springs, Michigan 49104-0380}                                    
\par \filbreak                                                                                     
  P.~Antonioli,                                                                                    
  G.~Bari,                                                                                         
  M.~Basile,                                                                                       
  L.~Bellagamba,                                                                                   
  D.~Boscherini,                                                                                   
  A.~Bruni,                                                                                        
  G.~Bruni,                                                                                        
  G.~Cara~Romeo,                                                                                   
  L.~Cifarelli,                                                                                    
  F.~Cindolo,                                                                                      
  A.~Contin,                                                                                       
  M.~Corradi,                                                                                      
  S.~De~Pasquale,                                                                                  
  P.~Giusti,                                                                                       
  G.~Iacobucci,                                                                                    
  A.~Margotti,                                                                                     
  R.~Nania,                                                                                        
  F.~Palmonari,                                                                                    
  A.~Pesci,                                                                                        
  G.~Sartorelli,                                                                                   
  A.~Zichichi  \\                                                                                  
  {\it University and INFN Bologna, Bologna, Italy}~$^{e}$
% ...
\newpage                                                                                           
$^{\    1}$ also affiliated with University College London \\                                      
$^{\    2}$ on leave of absence at University of                                                   
Erlangen-N\"urnberg, Germany\\                                                                     
% ...
                                                           %
                                                           %
\newpage   
                                                           %
                                                           %
\begin{tabular}[h]{rp{14cm}}                                                                       
$^{a}$ &  supported by the Natural Sciences and Engineering Research                               
          Council of Canada (NSERC) \\                                                             
$^{b}$ &  supported by the German Federal Ministry for Education and                               
          Research (BMBF), under contract numbers HZ1GUA 2, HZ1GUB 0, HZ1PDA 5, HZ1VFA 5\\         
$^{c}$ &  supported by the MINERVA Gesellschaft f\"ur Forschung GmbH, the                          
          Israel Science Foundation, the U.S.-Israel Binational Science                            
          Foundation and the Benozyio Center                                                       
          for High Energy Physics\\                                                                
$^{d}$ &  supported by the German-Israeli Foundation and the Israel Science                        
          Foundation\\                                                                             
$^{e}$ &  supported by the Italian National Institute for Nuclear Physics (INFN) \\                
$^{f}$ &  supported by the Japanese Ministry of Education, Culture,                                
          Sports, Science and Technology (MEXT) and its grants for                                 
          Scientific Research\\                                                                    
$^{g}$ &  supported by the Korean Ministry of Education and Korea Science                          
          and Engineering Foundation\\                                                             
$^{h}$ &  supported by the Netherlands Foundation for Research on Matter (FOM)\\                   
$^{i}$ &  supported by the Polish State Committee for Scientific Research,                         
          grant no. 620/E-77/SPUB-M/DESY/P-03/DZ 247/2000-2002\\                                   
$^{j}$ &  partially supported by the German Federal Ministry for Education                         
          and Research (BMBF)\\                                                                    
$^{k}$ &  supported by the Fund for Fundamental Research of Russian Ministry                       
          for Science and Edu\-cation and by the German Federal Ministry for                       
          Education and Research (BMBF)\\                                                          
$^{l}$ &  supported by the Spanish Ministry of Education and Science                               
          through funds provided by CICYT\\                                                        
$^{m}$ &  supported by the Particle Physics and Astronomy Research Council, UK\\                   
$^{n}$ &  supported by the US Department of Energy\\                                               
$^{o}$ &  supported by the US National Science Foundation\\                                        
$^{p}$ &  supported by the Polish State Committee for Scientific Research,                         
          grant no. 112/E-356/SPUB-M/DESY/P-03/DZ 301/2000-2002, 2 P03B 13922\\                    
$^{q}$ &  supported by the Polish State Committee for Scientific Research,                         
          grant no. 115/E-343/SPUB-M/DESY/P-03/DZ 121/2001-2002, 2 P03B 07022\\                    
\end{tabular}                                                                                      
                                                           %
                                                           %

\pagenumbering{arabic} 
\pagestyle{plain}
\section{Introduction}
\label{sec-int}

In $ep$ deep inelastic scattering (DIS) at HERA, final-state
hadrons are dominantly produced by interactions between virtual photons
and incoming protons.
Diffractive interactions, characterized by a large rapidity gap
in the distribution of the final-state hadrons,
have been observed and extensively studied at HERA
\cite{pl:b315:481,pl:b348:681,zfp:c68:569,zfp:c76:613,epj:c1:81,pl:b428:206,epj:c6:43,pl:b516:273,pr:d65:052001}.
The measurements of the diffractive DIS cross
sections~\cite{pl:b348:681,zfp:c68:569,zfp:c76:613,epj:c1:81,epj:c6:43}
have been quantified in terms of a diffractive structure function, $F_2^D$,
defined in analogy with the proton structure function, $F_2$.
The diffractive parton densities, determined from these measurements,
are dominated by gluons.
The diffractive process at HERA has often been considered to proceed through
the exchange of an object carrying the quantum numbers of the vacuum,
called the Pomeron ($\pom$). In the resolved-Pomeron model~\cite{pl:b152:256},
the exchanged Pomeron acts as a source of partons, one of which interacts
with the virtual photon. In an alternative view,
the diffractive process at HERA can be described by
the dissociation of the virtual photon into a $q{\bar q}$ or
$q{\bar q}g$ state which interacts with the proton by the exchange
of two gluons or, more generally, a gluon ladder
with the quantum numbers of the vacuum~\cite{pl:b406:171,epj:c11:111,zfp:c74:671}.

Charm production in diffractive DIS,
which has also been measured by the H1 and
ZEUS collaborations~\cite{pl:b520:191,pl:b545:244},
allows quantitative tests of the models due to the sensitivity
of charm production to gluon-initiated processes~\cite{proc:hera:1996:691}.
Calculations based on a gluon-dominated resolved Pomeron
predict a large charm rate in diffractive DIS~\cite{zfp:c70:89,hep-ph-9806340}.
In the two-gluon-exchange models, the rate from  the $q{\bar q}g$ state
is similar to that predicted by the resolved-Pomeron model, while the rate
from the $q{\bar q}$ state is lower.

In this analysis, charm production, tagged using $\dstar$ mesons, is studied in
diffractive interactions identified by the presence of a large rapidity gap
between the proton at high rapidities and the centrally-produced hadronic system.
The luminosity for the present measurement is about two times larger
than in the previous ZEUS study~\cite{pl:b545:244}.
The increase in luminosity and an improved rapidity acceptance in the proton
direction allow a more detailed comparison with
the model predictions in a wider kinematic range.
The open-charm contribution to the diffractive proton structure function
is measured for the first time.

\section{Experimental set-up}
\label{sec:expset}

The analysis was performed with data taken from 1998 to 2000, when HERA
collided electrons or positrons\footnote{Hereafter, both $e^+$ and $e^-$
are referred to as electrons, unless explicitly stated otherwise.}
with energy $E_e = 27.5\gev$ with protons
of energy $E_p = 920\gev$ yielding a centre-of-mass energy of $318\gev$.
The results are based on the sum of the $e^-p$
and $e^+p$ samples, corresponding to integrated luminosities of
$16.4~\pm~0.3\,$pb$^{-1}$ and $65.3~\pm~1.5\,$pb$^{-1}$, respectively.

\Zdetdesc
\Zctddesc\ZcoosysfnBeta

\Zcaldesc

In 1998-2000, the forward plug calorimeter (FPC)~\cite{fpcnim:a450:235}
was installed in the $20 \times 20$ cm${}^2$ beam hole of the FCAL,
with a small hole of radius $3.15$ cm in the centre to accommodate the
beam pipe. The FPC increased the forward calorimetric coverage 
by about 1 unit of pseudorapidity to $\eta\,\leq\,5$. The FPC consisted of
a lead--scintillator sandwich calorimeter divided longitudinally into
electromagnetic and hadronic sections that were read out separately by
wavelength-shifting  fibers and photomultipliers. The energy resolution,
as measured under test-beam conditions, was
$\sigma(E)/E = 0.41/\sqrt{E} \oplus 0.062$ and 
$\sigma(E)/E = 0.65/\sqrt{E} \oplus 0.06$ for electrons and pions,
respectively, with $E$ in GeV.

The position of electrons scattered at a small angle with respect to the 
electron beam direction was measured using the small-angle rear tracking 
detector (SRTD)~\cite{nim:a401:63}.
The luminosity was determined from the rate of the bremsstrahlung process
$ep \rightarrow e \gamma p$, where the photon was measured with a 
lead--scintillator calorimeter~\cite{desy-92-066,*zfp:c63:391,*acpp:b32:2025} 
located at $Z = -107\met$.

\section{Kinematics and reconstruction of variables}
\label{sec-kin}

The four-momenta $k$, $k'$ and $P$ label the incoming electron, outgoing
electron and the incoming proton, respectively, in DIS events:
$$e(k) + p(P) \rightarrow e(k') + \rm{anything}.$$
To describe the kinematics of DIS events, any two of the following invariants 
can be used:
$$Q^2 = -q^2 = -(k - k')^2;~ x = \frac{Q^2}{2P\cdot q};~ y = \frac{P\cdot q}{P\cdot k};~
W^2 = \frac{Q^2(1 - x)}{x},$$
where $Q^2$ is the negative square of the four-momentum $q$ carried by
the virtual photon, $x$ is the Bjorken scaling variable, $y$ is the
fraction of the electron energy transferred to the proton in its rest
frame, and $W$ is the centre-of-mass energy of the photon-proton system.
The scattered electron was identified using an algorithm based on
a neural network~\cite{nim:a365:508,*nim:a391:360}.
The hadronic final state was reconstructed using
combinations of CTD tracks and energy clusters measured in the CAL and
FPC to form energy-flow objects (EFOs) \cite{epj:c1:81,epj:c6:43,briskin:phd:1998}.
The kinematic variables were reconstructed using the double-angle
method~\cite{proc:hera:1991:23,*proc:hera:1991:43}.

To describe the diffractive process $ep \rightarrow e X p$,
where $X$ is the hadronic final state originating from the dissociation
of the virtual photon, two additional variables were used:
\begin{itemize}
\item{$x_\pomsub = \frac{Q^2 + M^2_X}{Q^2 + W^2}$,
where $M_X$ is the invariant mass of the system $X$. This variable is
the fraction of the incoming proton momentum carried by the diffractive exchange;
}
\item{$\beta = \frac{x}{x_\pomsub} = \frac{Q^2}{Q^2 + M^2_X}.$
In an interpretation in which partonic structure is ascribed to the
diffractive exchange, $\beta$ is the longitudinal momentum fraction of
the exchange that is carried by the struck quark.
}
\end{itemize}

The above expressions neglect the proton mass. The square of the
four-momentum transfer at the proton vertex, $t$, was not measured;
thus all results are implicitly integrated over this variable, which was
assumed to be zero in the expressions for $x_{\pomsub}$ and $\beta$.

The mass of the diffractive system $X$ was calculated from EFOs using:
$$M^2_X = \left( \sum_{i}E_i \right)^2 - \left(\sum_{i}P_{X,i}\right)^2 -
\left(\sum_{i}P_{Y,i}\right)^2 - \left(\sum_{i}P_{Z,i}\right)^2,$$
where the sum $i$ runs over the EFOs not associated with
the scattered electron.

The process studied in this paper is 
$ep \rightarrow e X p \rightarrow e (D^{*\pm} X') p$,
in which the system $X$ includes at least one $D^{*\pm}$ meson. The latter
was reconstructed using the mass-difference method~\cite{prl:35:1672}
in the decay channel 
$D^{*+} \rightarrow D^0 \pi_s^+$ followed by $D^0 \rightarrow K^- \pi^+ (+c.c.)$,
where $\pi_s$ indicates the ``slow'' pion. The fractional momentum of the
$D^{*\pm}$ in the photon-proton system is defined as
$$x(D^{*\pm}) = \frac{2|p^*(D^{*\pm})|}{W},$$
where $p^*(D^{*\pm})$ is the $D^{*\pm}$ momentum in the photon-proton
centre-of-mass frame.

\section{Models of diffractive charm production}
\label{sec-theory}

\begin{figure}[tbhp]
\begin{center}
\vspace*{-2.0cm}
\centerline{
\epsfysize=10cm
\epsffile{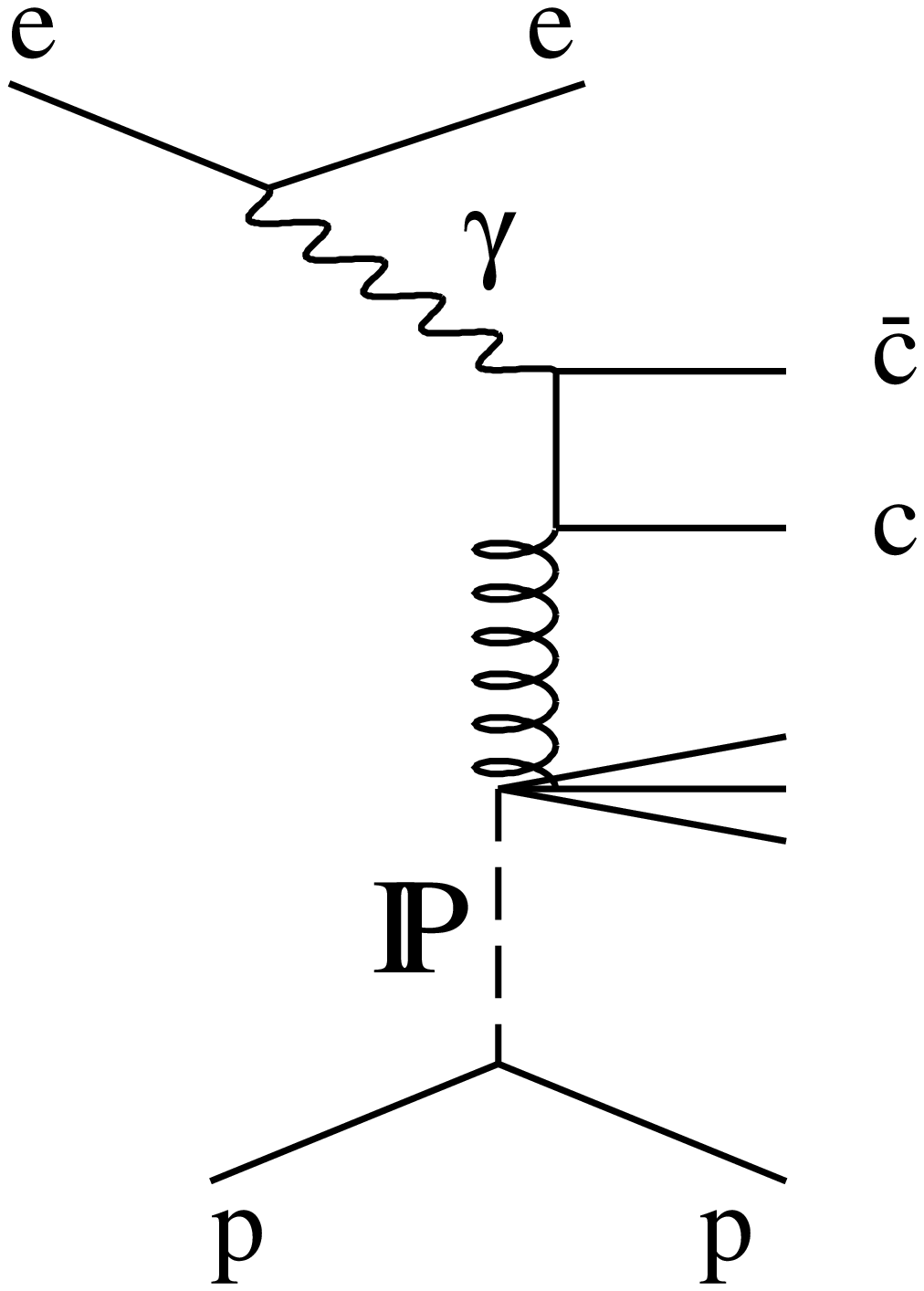}}
  \vskip-0.5cm
  \centerline{\Large\bf\kern7.1cm (a)\hfill}
  \vskip-0.5cm
\centerline{
\epsfysize=10cm
\hspace*{1.0cm}\epsffile{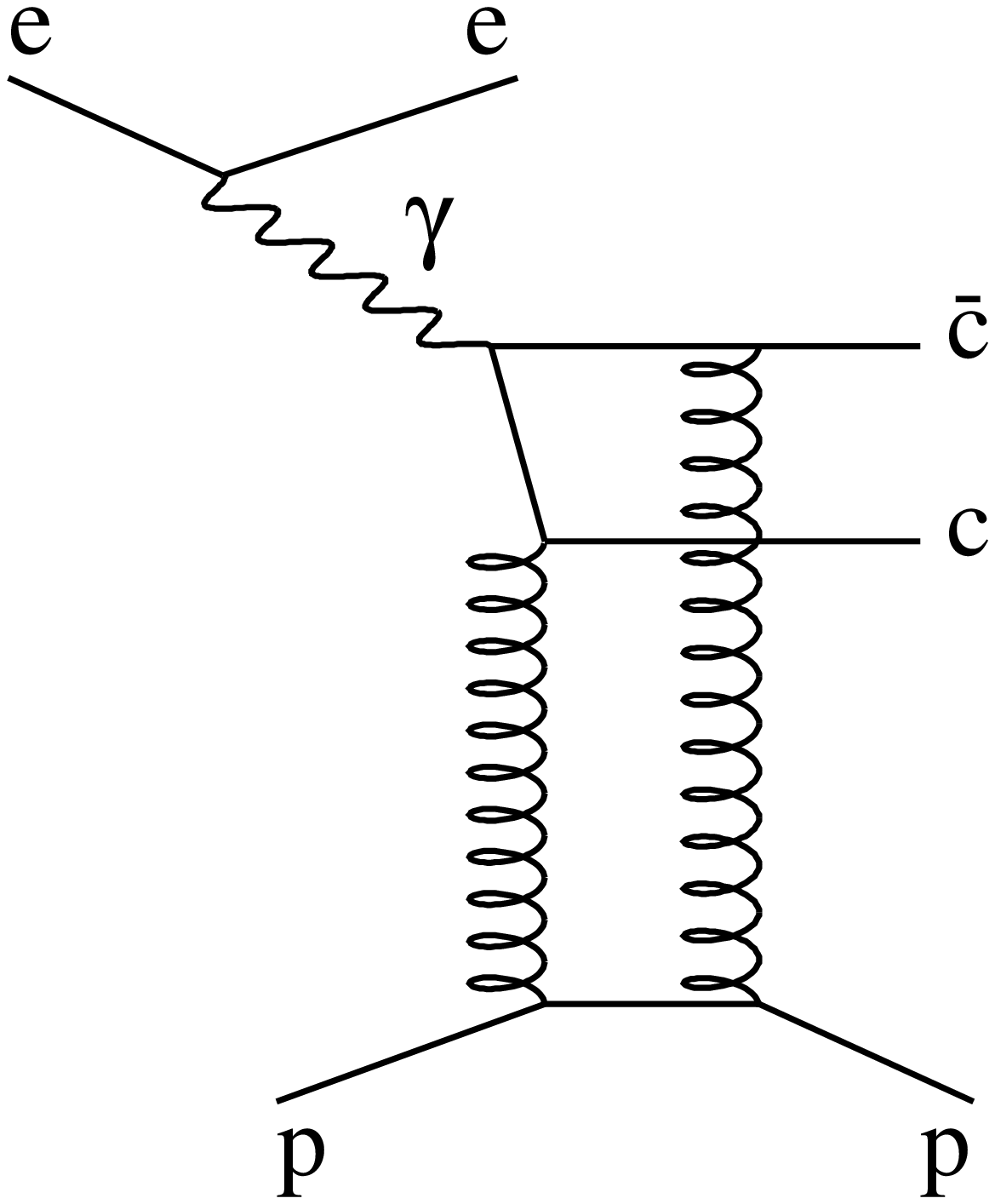}
\epsfysize=10cm
\hspace*{-2.5cm}\epsffile{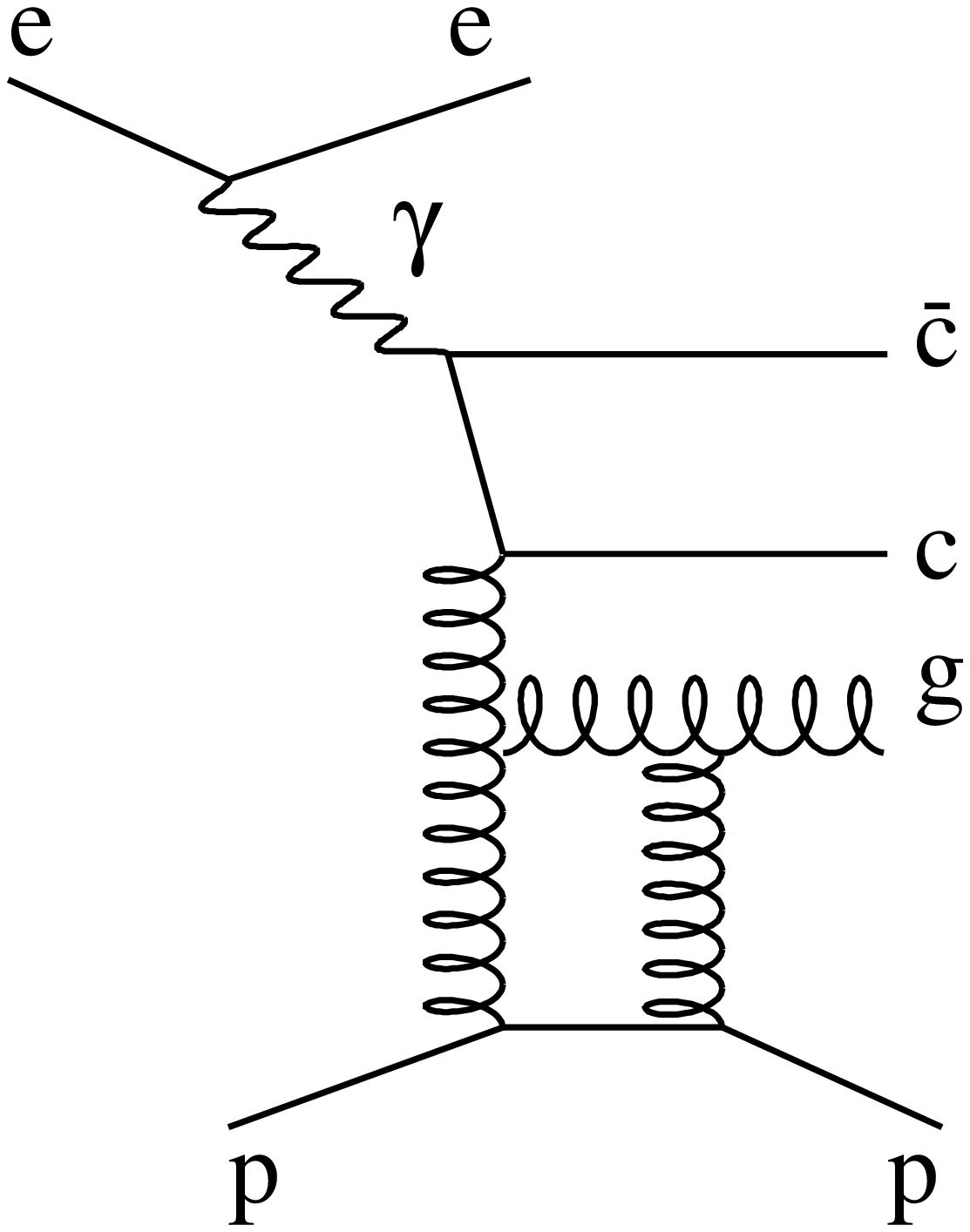}}
\end{center}
  \vskip-1.0cm
  \centerline{\Large\bf\kern4.6cm (b)\kern6.3cm (c)\hfill}
  \vskip0.5cm
\caption{Modelling charm production in diffractive $ep$ scattering:
(a) boson-gluon fusion in the resolved-Pomeron model,
(b) $c{\bar c}$ and (c) $c{\bar c}g$ states in the two-gluon-exchange model.
}
\label{fig-dgr}
\end{figure}

In the {\it resolved-Pomeron} model, proposed by Ingelman and
Schlein~\cite{pl:b152:256}, the exchanged Pomeron is assumed to be a object
with a partonic structure. The diffractive cross section factorises into a
Pomeron flux factor, describing the probability to find a Pomeron in the proton;
the Pomeron's parton density function (PDF), specifying the probability
to find a given parton in the Pomeron; and the interaction cross section
with the parton. Within this model, open charm is
produced in diffractive DIS via the boson-gluon-fusion (BGF) process,
where the virtual photon interacts with a gluon from the Pomeron
(Fig.~\ref{fig-dgr}a). The HERA measurements of the inclusive diffractive
differential cross sections
were found to be consistent with the resolved-Pomeron model
with a Pomeron structure dominated by gluons. For $\xpom > 0.01$,
an additional contribution from Reggeon exchanges, carrying the quantum numbers
of a $\rho$, $\omega$, $a$ or $f$ meson, was found to be sizeable~\cite{zfp:c76:613}.
A combined fit of the Pomeron parton densities to the H1 and ZEUS inclusive
diffractive DIS measurements~\cite{zfp:c76:613,zfp:c68:569,zfp:c70:391,epj:c1:81}
and to the ZEUS data on diffractive dijet photoproduction~\cite{pl:b356:129}
has been made by Alvero et al. (ACTW)~\cite{pr:d59:74022}.
The Pomeron flux factor was assumed to be of the Donnachie-Landshoff
form~\cite{pl:b191:309,*np:b303:634} and only data satisfying $\xpom < 0.01$
were used. To fit the Pomeron parton densities, five functional forms
(labelled A, B, C, D and SG) were used. It was found that
only gluon-dominated fits (B, D and SG) were able to describe
both the DIS and photoproduction data, while the quark-dominated fits (A and C)
underestimated the photoproduction data significantly. Therefore, only
the gluon-dominated fits are compared to the data in Section 8.
The fit results have been interfaced to the program HVQDIS~\cite{pr:d57:2806}
to calculate cross sections for diffractive charm production
in DIS~\cite{hep-ph-9806340}, both to leading and next-to-leading order (NLO)
in QCD. In this analysis, the ACTW NLO predictions were calculated setting
the charm-quark mass $m_c~=~1.45\gev$ and the renormalisation and factorisation
scales $\mu_R = \mu_F = \sqrt{Q^2 + 4m^2_c}$ as in~\cite{hep-ph-9806340}.
The Peterson fragmentation function (with $\epsilon = 0.035$~\cite{pr:d27:105})
was used for the charm decay. The probability for charm to fragment into a
$D^{*\pm}$ meson was set to $\fcds=0.235$~\cite{hep-ex-9912064}.

The {\it two-gluon-exchange} models consider fluctuations of the virtual photon
into $q{\bar q}$ or $q{\bar q}g$ colour dipoles that interact with the proton
via colour-singlet exchange; the simplest form of which is a pair
of gluons~\cite{pr:d12:163,*prl:34:1286,*pr:d14:246}.
The virtual-photon fluctuations into $c{\bar c}$ (Fig.~\ref{fig-dgr}b)
and $c{\bar c}g$ states (Fig.~\ref{fig-dgr}c)
can lead to diffractive open-charm production.
At high $\xpom$ values, quark exchanges are expected to become significant.
Thus, the two-gluon-exchange calculations are expected to be valid
only at low $\xpom$ values ($\xpom<0.01$). In recent
calculations~\cite{pl:b379:239,epj:c11:111,BJK:c24:555,pr:d59:014017},
the cross section for two-gluon exchange is related to the square of the
unintegrated gluon distribution of the proton which depends on the gluon
transverse momentum, $k_T$, relative to the proton direction.
In the ``saturation'' model
\cite{pr:d59:014017,proc:ringberg:1999:361n,*uproc:dis:2000:192},
the calculation of the $q{\bar q}g$ cross section is performed
under the assumption of strong $k_T$ ordering of the final-state partons,
which corresponds to $k_T^{(g)}\ll k_T^{(q,{\bar q})}$.
The parameters of the model were tuned to describe the total
photon-proton cross section measured at HERA. Alternatively, in the model of
Bartels et al.~\cite{pl:b379:239,epj:c11:111,BJK:c24:555},
configurations without strong $k_T$ ordering are included
in the $q{\bar q}g$ cross-section calculation and the minimum value for
the final-state-gluon transverse momentum, $k_{T,g}^{\rm cut}$, is a free
parameter. The sum of the $c{\bar c}$ and $c{\bar c}g$ contributions
in the saturation model and the model of Bartels et al. are hereafter referred
to as SATRAP and BJLW, respectively. Both the SATRAP and BJLW predictions were
calculated using the MC generator RAPGAP 2.08/06~\cite{cpc:86:147},
the proton PDF parameterisation GRV94HO~\cite{zfp:c67:433}, $m_c~=~1.45\gev$
and $\mu_R~=~\mu_F~=~\sqrt{p_{c,T}^2~+~4m_c^2}$, where $p_{c,T}$ is the
transverse momentum of the charm quark. Such scale form was used because
RAPGAP does not provide the form used for the ACTW predictions.
The probability for open charm to
fragment into a $D^{*\pm}$ meson was set to $\fcds=0.235$.
In the BJLW calculation of the $c\bar{c}g$ component, the value
of the parameter $k_{T,g}^{\rm cut}$ was set to $1.5\gev$~\cite{hep-ph-0204269}.

\section{Acceptance calculation}
\label{sec-mc}

To study trigger and selection efficiencies, two MC programs,
RAPGAP and RIDI 2.0~\cite{sovjnp:52:529,*proc:MC:1998:386},
were used to model the final states in the process
$ep \rightarrow e X p \rightarrow e (D^{*\pm} X') p$.

The RAPGAP generator was used in the resolved-Pomeron mode, in which charm
quarks are produced via the leading-order BGF process of Fig.~\ref{fig-dgr}a.
The higher-order QCD corrections were simulated using the colour-dipole model
implemented in ARIADNE 4.03~\cite{cpc:71:15}. The LUND string
model~\cite{prep:97:31} as implemented in JETSET 7.4~\cite{cpc:82:74}
was used for hadronisation. The charm-quark mass was set to the default
value of $1.5\gev$.
The diffractive sample was generated assuming a gluon-dominated
Pomeron, with a parameterisation from the H1 Collaboration
called ~``H1~fit~2''~\cite{pl:b520:191}. The Reggeon (meson) component of
the parameterisation was not used.

The RIDI generator is based on the two-gluon-exchange model developed by 
Ryskin~\cite{sovjnp:52:529,*proc:MC:1998:386}. To simulate the gluon momentum
density, the GRV94HO proton PDF parameterisation
was used. Final-state parton showers and hadronisation were simulated
using JETSET and the charm-quark mass was set to the default value of
$1.35\gev$. First-order radiative corrections were included in the simulation
although their effects were negligible. The $c\bar{c}$ and $c\bar{c}g$ components
were generated separately and later combined in the proportion $16\%:84\%$
which provided the best description of the $\beta$ distribution of the data.

The RAPGAP MC sample was used to evaluate the acceptance. Three MC samples
were used to estimate the model dependence of the acceptance corrections:
the RIDI MC sample, a sample generated with RAPGAP using parton showers as
implemented in LEPTO 6.1~\cite{cpc:101:108} to simulate the higher-order
QCD corrections, and a sample generated with RAPGAP using the Pomeron PDF
parameterisation ``H1~fit~3''~\cite{pl:b520:191}.

To estimate the non-diffractive DIS background and to
measure the ratio of diffractive to inclusive $\dstar$ production
(see \Sect{rat}), two MC generators were used:
RAPGAP in the non-diffractive mode for the nominal calculations
and HERWIG~6.301~\cite{cpc:67:465} as a systematic check. The RAPGAP
parameters used were the same as those used in the ZEUS measurement of the 
inclusive DIS $\dstar$ cross sections \cite{epj:c12:35}.
To generate charm production via the leading-order BGF process with HERWIG,
the CTEQ5L~\cite{epj:c12:375} proton PDF parameterisation
and $m_c=1.5\,$GeV were used. Hadronisation in HERWIG is simulated with
a cluster algorithm~\cite{np:b238:492}.

In this analysis, the final-state proton was not detected.
To estimate and subtract the contribution from the diffractive processes
where the proton dissociates into a system $N$,
$ep \rightarrow e X N \rightarrow e (D^{*\pm} X') N$, four MC generators
were used: DIFFVM~\cite{proc:mc:1998:396} for the nominal calculations
and RAPGAP, PHOJET~\cite{zfp:c66:203,*pr:d54:4246}
and EPSOFT 2.0~\cite{thesis:kasprzak:1994} for systematic checks.
The DIFFVM MC program provides a detailed description of the
proton-dissociative final state. The mass spectrum, $M_N$, of the system $N$
is generated as a superposition of $N^{*+}$ resonances and a
continuum having the form $d\sigma/dM_N^2 \propto M_N^{-2(1+\epsilon)}$.
The default parameter value $\epsilon = 0.0808$~\cite{pl:b296:227} was used.
In the RAPGAP simulation of proton dissociation, the proton splits into a
quark and di-quark and the Pomeron is assumed to couple only to the single quark.
The $M_N$ spectrum follows a $1/M_N$ distribution. In PHOJET, $M_N$ is
calculated from the triple-Pomeron kinematics~\cite{zfp:c66:203,*pr:d54:4246}
and an approximation of the low-mass-resonance structure. In EPSOFT,
the $M_N$-spectrum generation relies on a parameterisation of the
$pp \rightarrow pN$ data.

The generated events were passed through the GEANT-based
\cite{tech:cern-dd-ee-84-1} simulation of the ZEUS detector and trigger.
They were reconstructed by the same program chain as the data.

\section{Event selection and $\dspm$ reconstruction}
\label{sec-sel}

\subsection{Trigger and DIS selection}

Events were selected online with a three-level 
trigger~\cite{zeus:1993:bluebook,uproc:chep:1992:222}.
At the first level, events with an electron candidate
in the EMC sections of RCAL or BCAL were selected~\cite{nim:a355:278}.
In the latter case, a coincidence with a track originating 
at the nominal interaction point was required.
At the second level, the non-$ep$ background was further reduced
by removing events with CAL timing inconsistent with an $ep$
interaction. At the third level, events were fully reconstructed
and selected by requiring a coincidence of a scattered-electron
candidate found within the CAL and a $\dstar$ candidate reconstructed
in the nominal decay mode using charged tracks measured by the CTD.
The requirements were similar to, but looser than, the offline
cuts described below. The efficiency of the online $\dstar$
reconstruction, determined relative to an inclusive DIS trigger,
was above $95\%$.

The following criteria were applied offline to select DIS events:
\begin{itemize}
\item an electron with energy above 10 GeV;
\item the impact point of the scattered electron on the RCAL lies
      outside the region 26$\times$14 cm$^2$ centred on the beamline;
\item 40 $< \delta <$ 65 GeV, where $\delta = \sum_i (E_i-P_{Z,i})$
      and the sum runs over the EFOs from the hadronic system and
      the energy deposited by the identified electron;
\item a vertex position $|Z_{\rm vtx}|<$ 50 cm.
\end{itemize}
The events were restricted to the kinematic region
$1.5<Q^2<200\gev^2$ and $0.02<y<0.7$.

\subsection{$\mathbf{\dstar}$ reconstruction}
\label{sec-drec}

Charged tracks with $p_T > 0.12\gev$ and $|\eta| < 1.75$ were selected.
Only tracks assigned to the primary event vertex and
with hits in at least three superlayers of the CTD were considered.
Two oppositely charged tracks, each with $p_T > 0.5\gev$,
were combined to form a $D^0$ candidate. The tracks were alternately
assigned the mass of a charged kaon and a charged pion
and the invariant mass of the track pair, $M(K \pi)$, was calculated.
Only $D^0$ candidates that satisfy $1.81 < M(K \pi) < 1.92\gev$ were kept.
Any additional track, with $p_T > 0.12\gev$ and charge opposite to that
of the kaon track, was assigned the pion mass and combined with
the $D^0$ candidate to form a $\dstar$ candidate with invariant
mass $M(K\pi\pi_s)$. The $\dspm$ candidates were required to have
$p_T(D^{*})>1.5\gev$ and $|\eta(D^{*})|<1.5$.

In the distribution of the mass difference, $\Delta M = M(K\pi\pi_s) - M(K\pi)$,
for selected $\dstar$ candidates, a clear signal at the nominal
value of $M(\dspm)-M(\dz)$ was observed (not shown).
The combinatorial background under this signal was estimated from
the mass-difference distribution for wrong-charge combinations,
in which both tracks forming the $D^0$ candidates have the same charge
and the third track has the opposite charge. The number of reconstructed
$D^{*\pm}$ mesons was determined by subtracting
the wrong-charge $\Delta M$ distribution after normalising it
to the $\Delta M$ distribution of \dspm\ candidates with the appropriate
charges in the range $\,0.15 < \Delta M < 0.17\gev$. The subtraction,
performed in the range $\,0.1435 < \Delta M < 0.1475\gev$, yielded an inclusive
signal of $4976\pm 103$ $D^{*\pm}$ mesons.

\subsection{Selection of diffractive events}
\label{sec:difsel}

\begin{figure}[tbhp]
\begin{center}
\hspace*{0.5cm}\includegraphics[height=160mm]{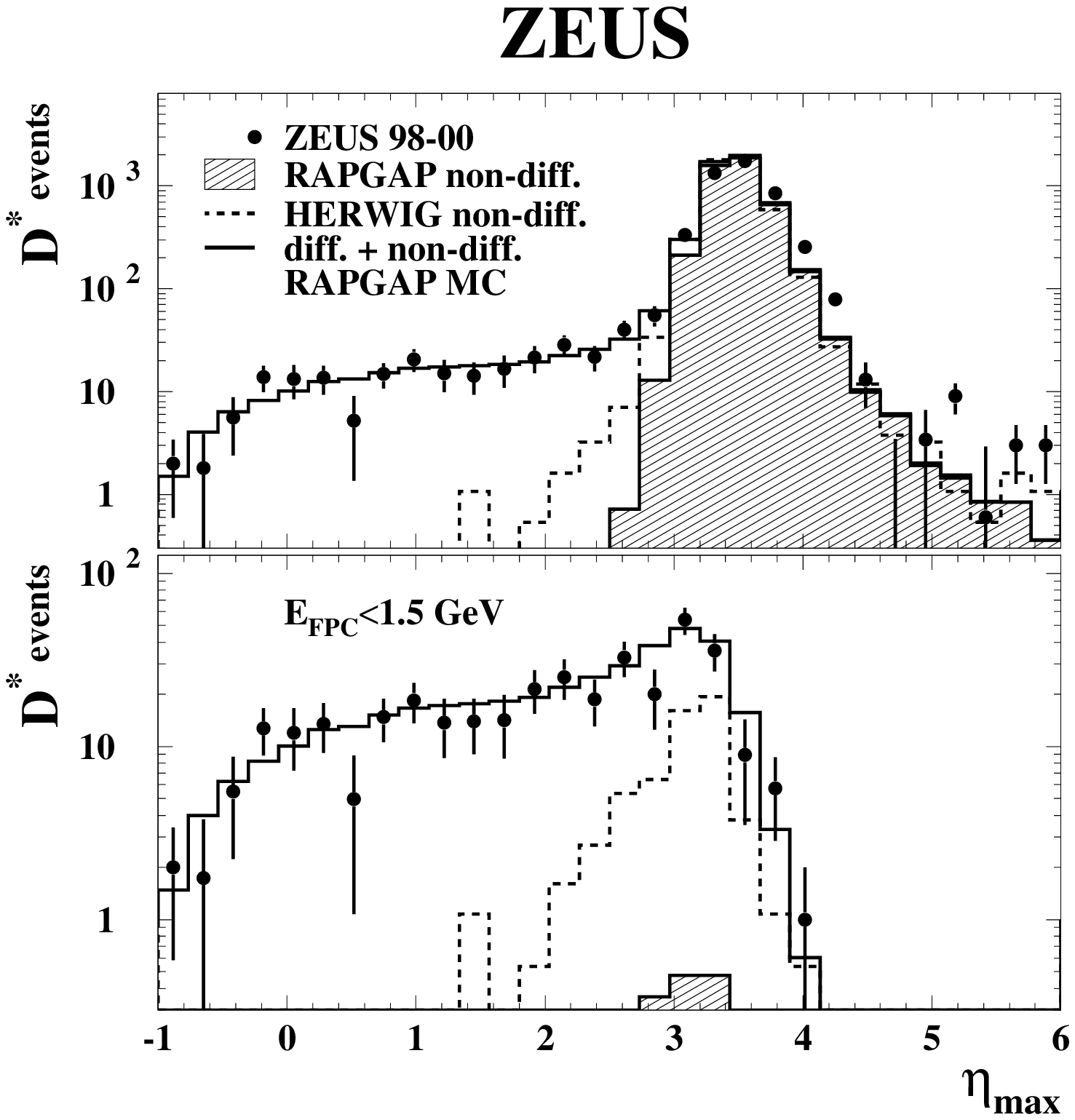}
\end{center}
  \vskip-13.7cm
  \centerline{\Large\bf\kern13.3cm (a)\hfill}
  \vskip5.6cm
  \centerline{\Large\bf\kern13.3cm (b)\hfill}
  \vskip6.5cm
\caption{
Numbers of reconstructed $D^{*\pm}$ mesons (dots)
as a function of $\eta_{\rm max}$
for DIS events with
(a) any $E_{\rm FPC}$ values and
(b) $E_{\rm FPC} < 1.5\gev$.
The solid histogram shows the sum of the non-diffractive
RAPGAP MC (hatched area)
and the diffractive RAPGAP MC. The sum was
normalised to have the same area as the data.
The dashed histogram shows the non-diffractive HERWIG MC.
}
\label{fig-fig1}
\end{figure}

Diffractive events are characterised by the presence of a large rapidity
gap between the proton at high rapidities and the centrally-produced 
hadronic system. To select such events,
the following two requirements were applied:
\begin{itemize}
\item $E_{\rm FPC}<1.5\,$GeV, where $E_{\rm FPC}$ is the energy
deposited in the FPC;
\item $\eta_{\rm max}<3$, where $\eta_{\rm max}$
is the pseudorapidity of the most-forward EFO
measured without using FPC information and with energy above $400\mev$.
\end{itemize}
This selection is illustrated in Fig.~\ref{fig-fig1}, where
the distribution of $\eta_{\rm max}$ is shown for $D^{*\pm}$ mesons
obtained after the wrong-charge-background subtraction.
The data are compared to the $\eta_{\rm max}$ distributions of the
non-diffractive RAPGAP and HERWIG MC samples and to the sum of
the non-diffractive and diffractive RAPGAP MC. In Fig.~\ref{fig-fig1}a,
the distributions are shown for events with any  $E_{\rm FPC}$ value.
The large peak at $\eta_{\rm max} \sim 3.5$ corresponds to non-diffractive
events in which the proton remnant deposits energy around the beam direction.
On the low side of the peak, the contribution from non-diffractive
interactions exhibits an exponential fall-off, leaving an excess at low
values of $\eta_{\rm max}$ which is populated predominantly by diffractive
events. Figure~\ref{fig-fig1}b shows that the requirement
$E_{\rm FPC}<1.5\,$GeV strongly suppresses the contribution from
non-diffractive interactions. Requiring $\eta_{\rm max}<3$ in addition
reduces the remaining non-diffractive background and ensures
a gap of at least two units of pseudorapidity with respect to the edge of
the forward calorimetric coverage (see Section~\ref{sec:expset}).

The selected events were analysed in terms of the diffractive variables
$x_\pomsub$, $\beta$ and $M_X$. To account for the restriction imposed by
the $\eta_{\rm max}<3$ requirement, a cut of $x_\pomsub < 0.035$ was applied.
In addition, a cut of $\beta < 0.8$ was also used because diffractive
charm production in DIS is strongly suppressed at large $\beta$ values due
the dominant contribution of events with small $Q^2$ and large $M_X$ values.

\begin{figure}[tbhp]
\begin{center}
\includegraphics[height=160mm]{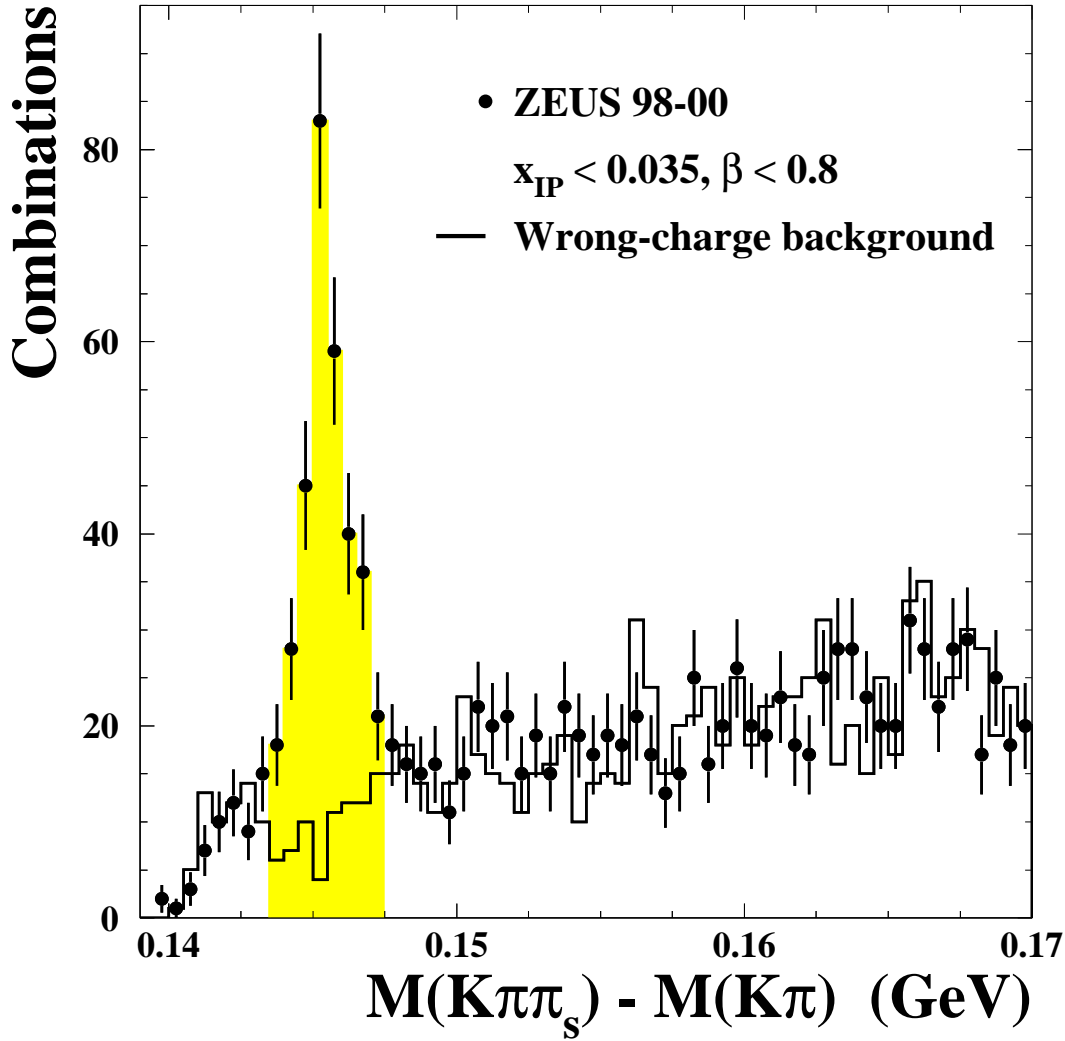}
\end{center}
\caption{
The distribution of the mass difference,
$\Delta M=M(K \pi \pi_s)-M(K \pi)$, for
\dspm\ candidates (dots) in events with
$\eta_{\rm max} < 3$, $E_{\rm FPC}<1.5\,$GeV,
$x_\pomsub < 0.035$ and $\beta < 0.8$.
The histogram
shows the $\Delta M$ distribution for wrong-charge combinations.
Only \dspm\ candidates from the shaded band
were used for the differential cross-section measurements.
}\label{fig-fig2}
\end{figure}

Figure~\ref{fig-fig2} shows the $\Delta M$ distribution after the above cuts.
The number of $D^{*\pm}$ after the wrong-charge-background subtraction
is $253 \pm 21$.

\begin{figure}[tbhp]
\begin{center}
\includegraphics[height=160mm]{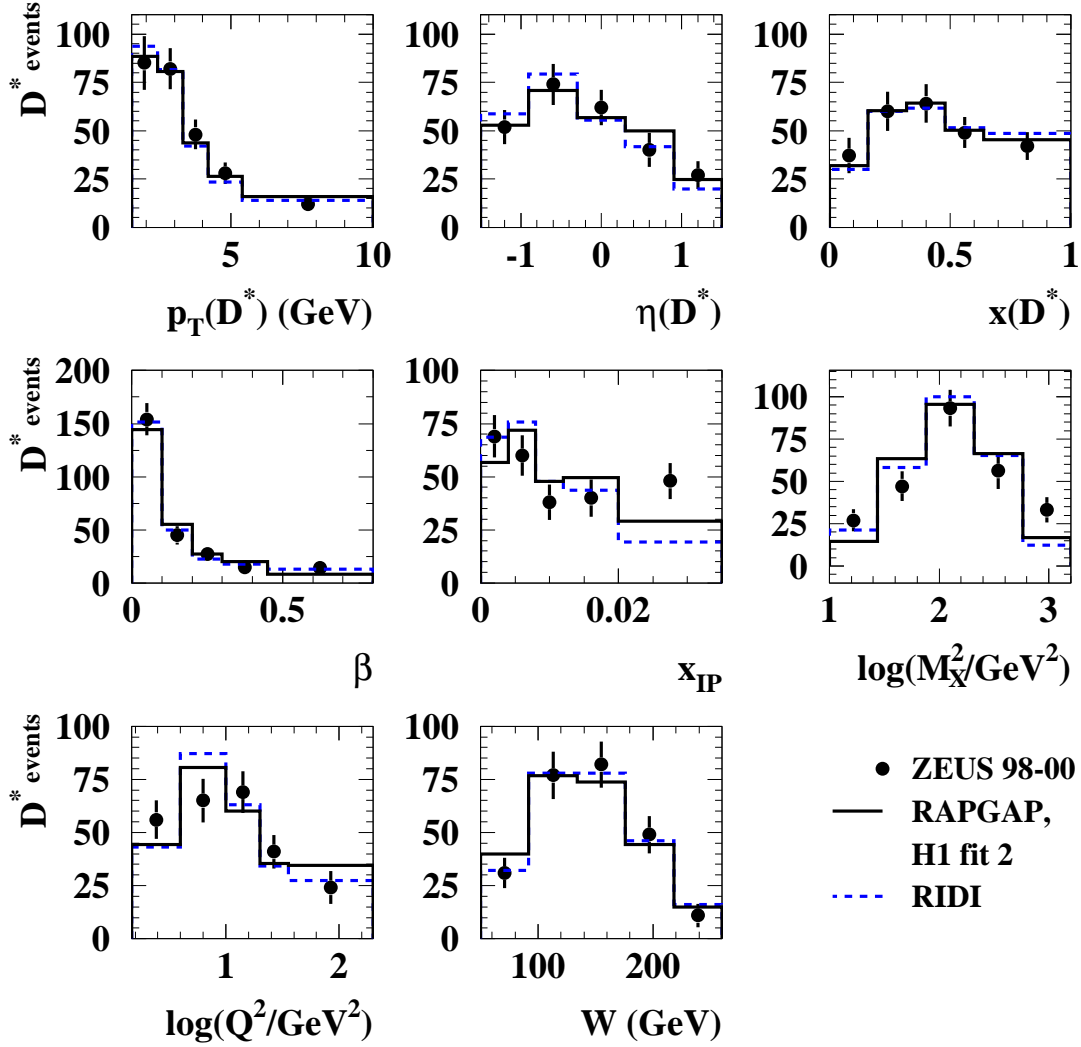}
\end{center}
\caption{
Numbers of reconstructed $D^{*\pm}$ mesons (dots) in bins of
$p_T(D^{*\pm})$, $\eta(D^{*\pm})$, $x(D^{*\pm})$, $\beta$, $x_\pomsub$,
$\log(M_X^2)$, $\log(Q^2)$ and $W$.
The RAPGAP (solid histogram) and the mixed $\ccb$ and $\ccb g$ RIDI
(dashed histogram) MC samples, normalized to the data,
are shown for comparison.
}\label{fig-fig4}
\end{figure}

Figure~\ref{fig-fig4} shows the number of reconstructed $D^{*\pm}$ mesons
in bins of the variables $p_T(D^{*\pm})$, $\eta(D^{*\pm})$, $x(D^{*\pm})$,
$\beta$, $x_\pomsub$, $\log (M_X^2)$, $\log (Q^2)$ and $W$. The data are compared to
the diffractive RAPGAP and RIDI simulations (normalised to the data).
Both simulations reproduce the shapes of the data.

\subsection{Subtraction of the proton-dissociative contribution}
\label{sec-pdis}

\begin{figure}[tbhp]
\begin{center}
\includegraphics[height=160mm]{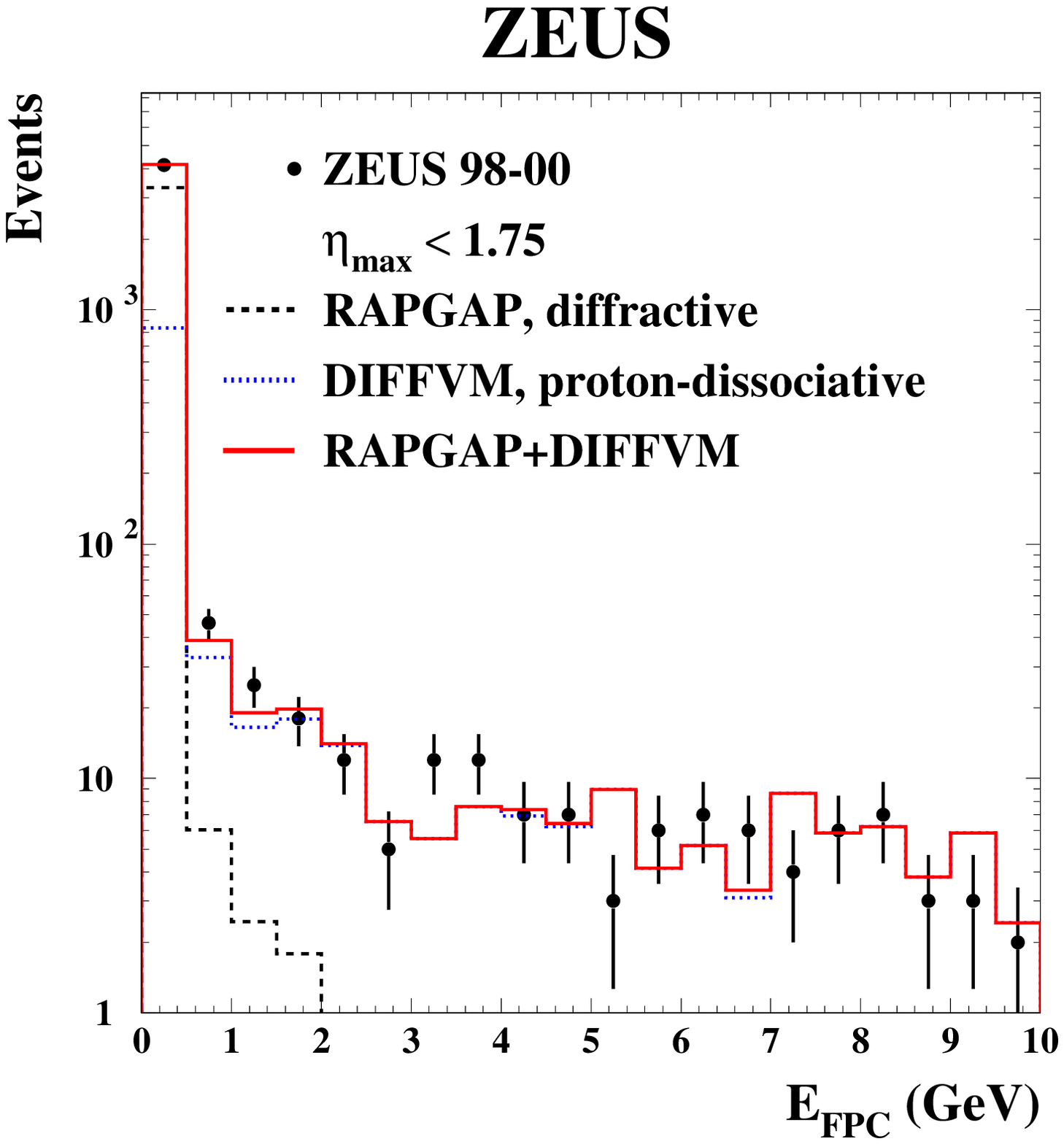}
\end{center}
\caption{The measured energy in the FPC for events
with $\eta_{\rm max} < 1.75$
(dots). The dashed histogram is the single-diffractive
RAPGAP MC sample and the dotted
histogram is the proton-dissociative DIFFVM MC sample. The solid histogram
is the sum of both diffractive and proton-dissociative MC samples
normalised to the data. }
\label{fig-fig3}
\end{figure}

Diffractive events with proton dissociation can pass the
$E_{\rm FPC}<1.5\,$GeV and $\eta_{\rm max}<3$ requirements
if the major part of the proton-dissociative system
escapes undetected down the forward beampipe.
The proton-dissociative contribution
was determined from the distribution of $E_{\rm FPC}$
for events selected with relaxed $\dstar$ reconstruction cuts
and without cutting on $E_{\rm FPC}$.
To ensure a gap of at least two units of pseudorapidity
between the proton-dissociative system, tagged by the FPC, and
the system $X$, a requirement of $\eta_{\rm max} < 1.75$ was applied.
Figure~\ref{fig-fig3} compares the $E_{\rm FPC}$ distribution for these
events to the distributions of the diffractive RAPGAP and proton-dissociative
DIFFVM MC samples. The MC samples were combined in the proportion
providing the best description of the $E_{\rm FPC}$ distribution,
and their sum was normalised to the data.
Using the normalisation factors obtained for the two MC samples,
the proton-dissociative contribution was calculated for the nominal
diffractive selection described in Section~\ref{sec:difsel}.
The proton-dissociative contribution was determined 
to be $16$\% with negligible statistical uncertainty;
the systematic uncertainty was obtained as follows,
where the effects of each source are shown in parentheses:
\begin{itemize}
\item{ the parameter $b$, regulating the shape
of the $M_N$ continuum distribution in the DIFFVM MC simulation,
was varied between $0.7$ and $1.5$ ($^{+ 3.7}_{- 3.0}$\%);}
\item{ uncertainties in the low-mass resonance structure and other details
of the simulation of the proton-dissociative system were estimated
by using the PHOJET, RAPGAP and EPSOFT MC generators ($^{+ 1.6}_{- 0.9}\%$);}
\item{ a shift of $\pm 10$\% due to the FPC energy-scale uncertainty
($^{+ 0.5}_{-0.1}$\%); }
\item{ a larger area, including the FPC and neighbouring FCAL towers,
was used to tag the proton-dissociative system ($-2.7\%$).
This check is sensitive to the high-$M_N$ proton-dissociative
contribution and to details of the FPC and FCAL simulation.}
\end{itemize}

These systematic uncertainties were added in quadrature separately for the 
positive and negative variations
to determine the overall systematic uncertainty of $\pm 4.1$\%.
The proton-dissociative contribution of $(16 \pm 4)\%$ was assumed
to be independent of all kinematic variables
and was subtracted from all measured cross sections.

\section{Systematic uncertainties}
\label{sec-syst}

The systematic uncertainties of the measured cross sections were determined
by changing the selection cuts or the analysis procedure in turn and repeating the
extraction of the cross sections~\cite{thesis:vlasov:2003}.
The major sources of the systematic uncertainty were as the follows, where
effects on the integrated cross section are shown in parentheses:

\begin{itemize}
\item the selection of inclusive DIS events ($^{+2.3}_{-3.3}$\%).
Variations were made in 
the cut on the scattered-electron energy, the RCAL box cut, the $\delta$
cut and the vertex-position cut. In addition, both $Q^2$ and $y$ were
determined using the $e\Sigma$ method~\cite{nim:a426:583} rather
than using the $DA$ method;
\item the selection of $\dstar$ candidates and background estimation
 ($^{+4.5}_{-3.7}\%$).
The minimum transverse momentum for the $K$ and $\pi$ candidates
was raised and lowered by $25$ MeV. For the slow pion, $\pi_s$,
the minimum transverse momentum was raised and lowered by $10$ MeV.
The signal region for $M(D^0)$ was loosened to $1.80 < M(D^0) < 1.93$ GeV
and that of the $\Delta M$ distribution was widened to
$0.143 < \Delta M < 0.148$ GeV. The $\Delta M$ background-normalisation
region was varied by $5\,$MeV;
\item the selection of diffractive events ($^{+3.9}_{-1.4}$\%).
The requirements on $\etam$
and $E_{\rm FPC}$ were varied by $\pm 0.2$ units and $\pm 0.5\,$GeV,
respectively;
\item a shift of $\pm 3$\% due to the CAL energy-scale uncertainty
($^{+ 0.7}_{-0.3}$\%);
\item a shift of $\pm 10$\% due to the FPC energy-scale uncertainty
($^{+ 0.2}_{-0.3}$\%);
\item the model dependence of the non-diffractive contribution ($-6.6$\%).
This uncertainty was estimated using the HERWIG sample;
\item the model dependence of the acceptance corrections ($^{+1.6}_{-7.4}\%$).
This uncertainty was estimated using the RIDI MC sample, the RAPGAP
sample generated with the LEPTO parton showers and the RAPGAP
sample generated with the ``H1~fit~3'' parameterisation
of the Pomeron structure function.
\end{itemize}

These systematic uncertainties were added in quadrature separately for the 
positive and negative variations to determine the overall systematic
uncertainty of $^{+6.6}_{-11.2}$\%. These estimates were also made
in each bin in which the differential cross sections were measured.

The normalisation uncertainties in the luminosity measurement ($\pm2.2\%$) and
the $\dstar$ and $D^0$ branching ratios ($\pm2.5\%$~\cite{pr:d55:10001})
were not included in the systematic uncertainty.  The uncertainty
arising from the subtraction of the proton-dissociative background,
quoted separately, is $\pm4.1\%/0.84 = \pm4.9\%$.

\section{Results}
\label{sec:results}
\subsection{Cross sections}
\label{sec-xs}

The differential $D^{*\pm}$ cross sections for any given variable $\xi$
were determined using:
$$\frac{d\sigma}{d\xi} = \frac{N(D^*)~(1 - f_{pd})}{A~\mathcal{L}~B~\Delta \xi},$$
where $N(D^*)$ is the number of $D^{*\pm}$ mesons in a bin of width
$\Delta \xi$, $A$ is the acceptance for that bin,
$\mathcal{L}$ is the integrated luminosity, $B$ is the product of the
$D^{*+} \rightarrow D^0 \pi_s^+$ and $D^0 \rightarrow K^- \pi^+$
branching ratios ($0.0257$~\cite{pr:d55:10001}),
and $f_{pd}$ ($0.16$) is the fraction of the
proton-dissociative background discussed in \Sect{pdis}.

Using the overall acceptance of $19.4\%$, the cross section for
diffractive $\dstar$ production in the kinematic region
$1.5<Q^2<200\gev^2$, $0.02<y<0.7$, $x_\pomsub<0.035$, $\beta<0.8$,
$p_T(D^{*\pm})>1.5\gev$ and $|\eta(D^{*\pm})| < 1.5$ is
$$\sigma_{ep \rightarrow e D^{*\pm} X' p} =
 521 \pm 43({\rm stat.})^{+34}_{-58}({\rm syst.}){\pm 26}({\rm p.diss.})\,{\rm pb},$$
where the last uncertainty arises from the subtraction
of the proton-dissociative background\footnote{
The diffractive $D^{*\pm}$ cross section was also calculated in the kinematic
regions in which previous measurements~\cite{pl:b520:191,pl:b545:244}
were reported and was found to be consistent.}.

In the case of Reggeon exchanges, open charm can be produced
in the BGF process if the exchanged-meson PDF contains gluons.
The Reggeon contribution to diffractive $D^{*\pm}$ production
in the measured kinematic range was estimated to be less
than $6\%$ using RAPGAP with the Pomeron and meson PDF parameterisations
``H1 fit 2'' or ``H1 fit 3''. The contribution is less than $0.5\%$ for $\xpom<0.01$;
it increases with $\xpom$, contributing about $12\%$ in the last bin.
The Reggeon contribution, which is smaller than the statistical uncertainty
of the measurement, was neglected.

\begin{figure}[tbhp]
\begin{center}
\includegraphics[height=160mm]{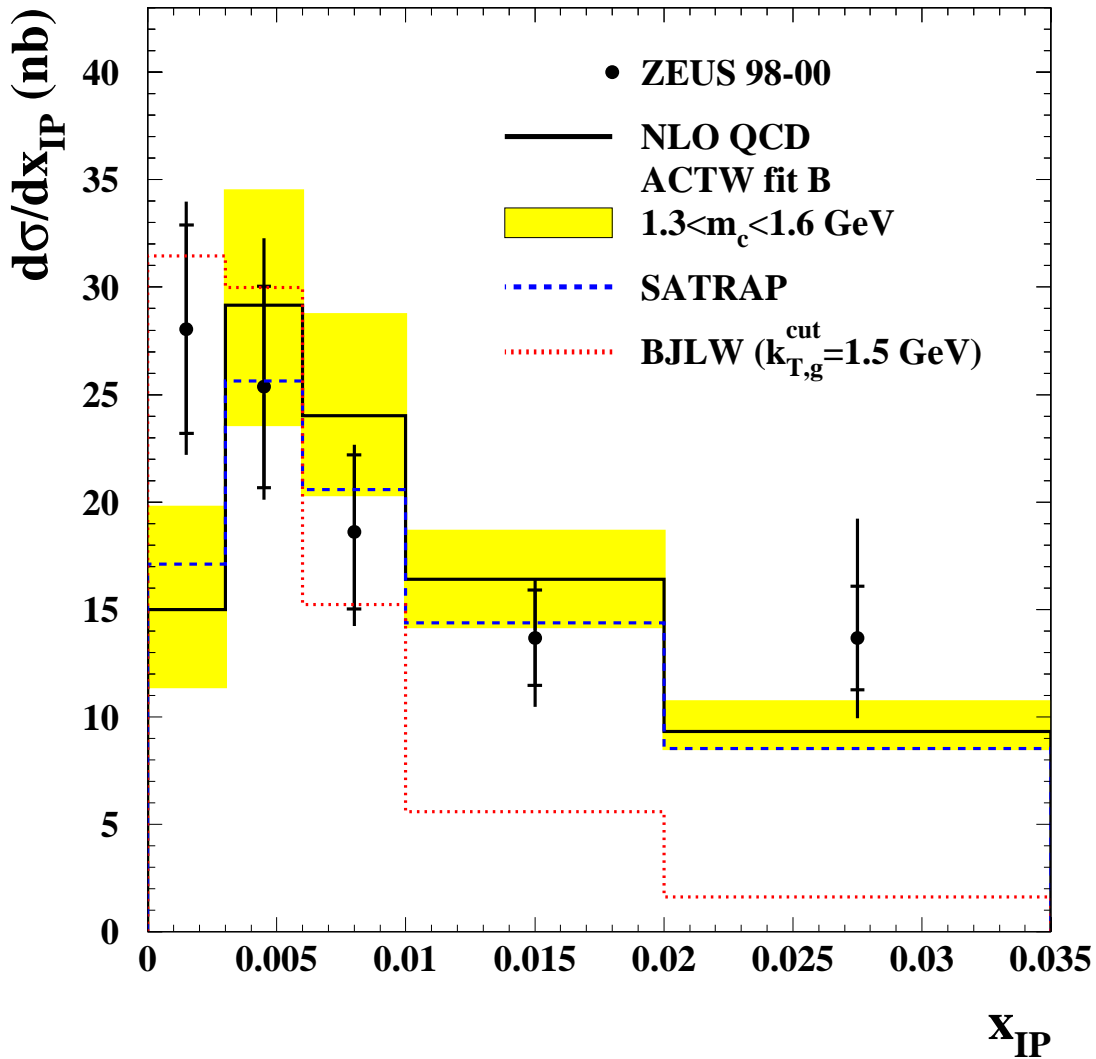}
\end{center}
\caption{Differential cross-section $\diff\sigma/\diff x_\pomsub$ for
diffractive $D^{*\pm}$ production
for the data (dots) compared with the
ACTW NLO (solid histogram), SATRAP (dashed histogram)
and BJLW (dotted histogram) predictions.
The shaded area shows the effect of
varying the charm-quark mass in the ACTW NLO prediction.
The inner error bars indicate the
statistical uncertainties, while the outer ones correspond to statistical and
systematic uncertainties added in quadrature.
The overall normalisation uncertainties arising from
the luminosity measurement ($\pm2.2\%$), from the $\dstar$ and $D^0$ branching
ratios ($\pm2.5\%$) and from the proton-dissociative background
subtraction ($\pm4.9\%$) are not indicated.
}\label{fig-fig5}
\end{figure}

Figure~\ref{fig-fig5} (Table~\ref{tab-xsecxp}) shows the differential
cross section as a function of $x_\pomsub$. The data are compared with
the ACTW NLO predictions, calculated with the gluon-dominated fit B,
the SATRAP predictions and the BJLW predictions.
All three models agree with the data within experimental uncertainties
below $x_\pomsub=0.01$. For  larger $x_\pomsub$ values,
the ACTW and SATRAP models agree with the data whereas
the BJLW prediction underestimates the measured cross sections
as expected (see \Sect{theory}).

\begin{figure}[tbhp]
\begin{center}
\includegraphics[height=160mm]{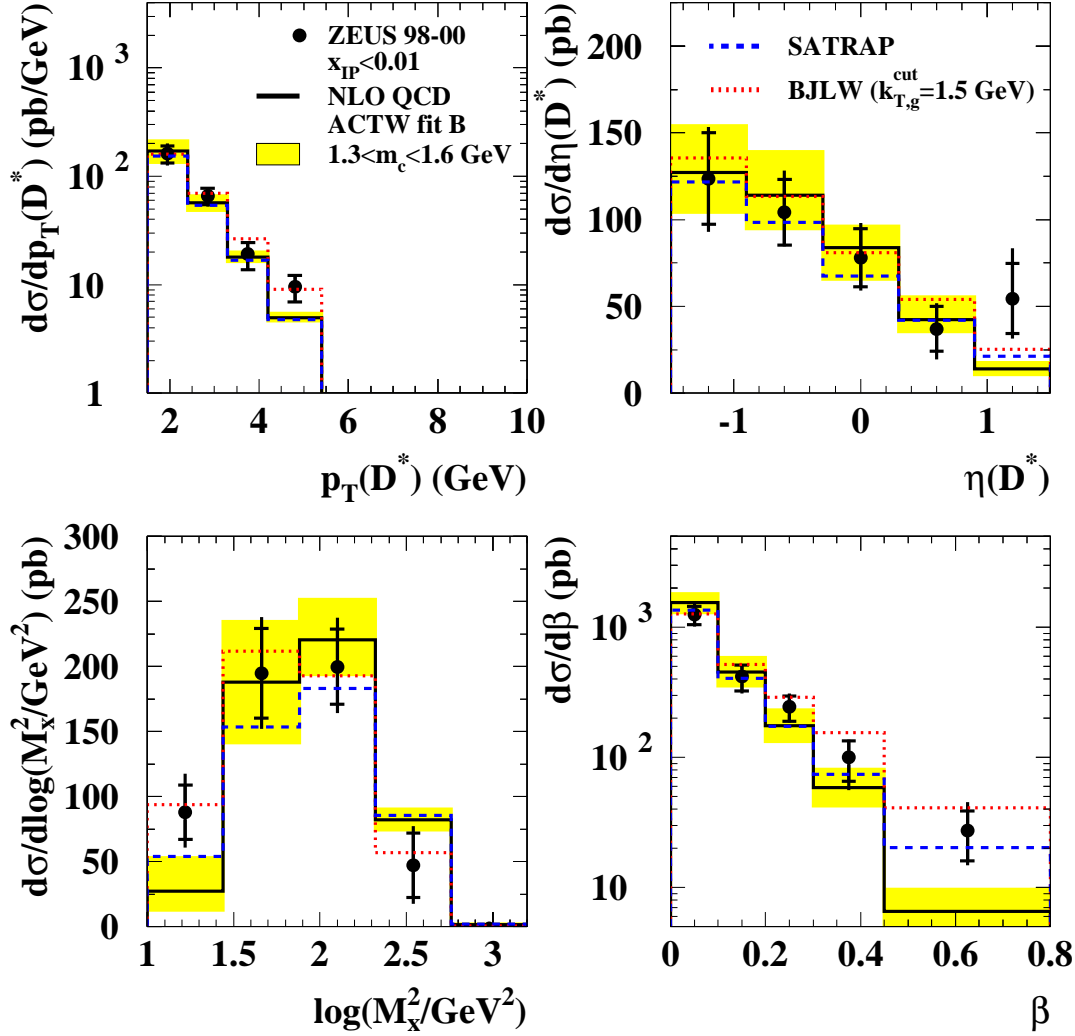}
\end{center}
\caption{Differential cross sections for diffractive $D^{*\pm}$ production
with $x_\pomsub <0.01$ for the data (dots) compared with the
ACTW NLO (solid histogram), SATRAP (dashed histogram)
and BJLW (dotted histogram) predictions.
The shaded area shows the effect of
varying the charm quark-mass in the ACTW NLO prediction.
The cross sections are
shown as a function of $p_T(D^{*\pm})$, $\eta(D^{*\pm})$, $\log(M_X^2)$
and $\beta$.
The inner error bars indicate the
statistical uncertainties, while the outer ones correspond to
statistical and systematic
uncertainties added in quadrature.
The overall normalisation uncertainties arising from
the luminosity measurement ($\pm2.2\%$), from the $\dstar$ and $D^0$ branching
ratios ($\pm2.5\%$) and from the proton-dissociative background
subtraction ($\pm4.9\%$) are not indicated.
}\label{fig-fig7}
\end{figure}

\begin{figure}[tbhp]
\begin{center}
\includegraphics[height=160mm]{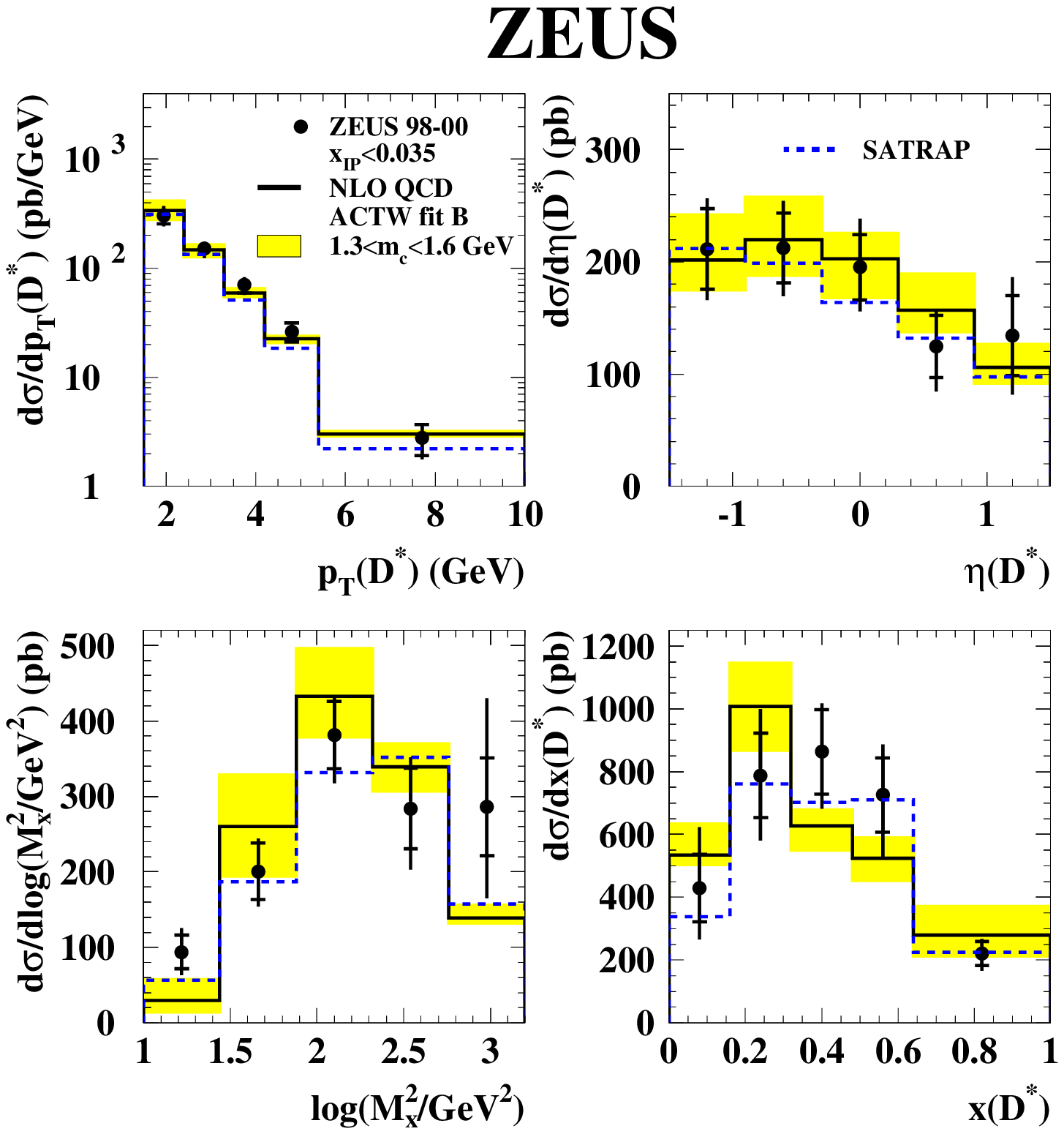}
\end{center}
\caption{Differential cross sections for diffractive $D^{*\pm}$ production
with $x_\pomsub <0.035$ for the data (dots) compared with the
ACTW NLO (solid histogram) and SATRAP (dashed histogram) predictions.
The shaded area shows the effect of
varying the charm-quark mass in the ACTW NLO prediction.
The cross sections are
shown as a function of $p_T(D^{*\pm})$, $\eta(D^{*\pm})$, $\log(M_X^2)$ and $x(D^{*\pm})$.
The inner error bars indicate the
statistical uncertainties, while the outer ones correspond to
statistical and systematic
uncertainties added in quadrature.
The overall normalisation uncertainties arising from
the luminosity measurement ($\pm2.2\%$), from the $\dstar$ and $D^0$ branching
ratios ($\pm2.5\%$) and from the proton-dissociative background
subtraction ($\pm4.9\%$) are not indicated.
}\label{fig-fig6a}
\end{figure}

\begin{figure}[tbhp]
\begin{center}
\includegraphics[height=160mm]{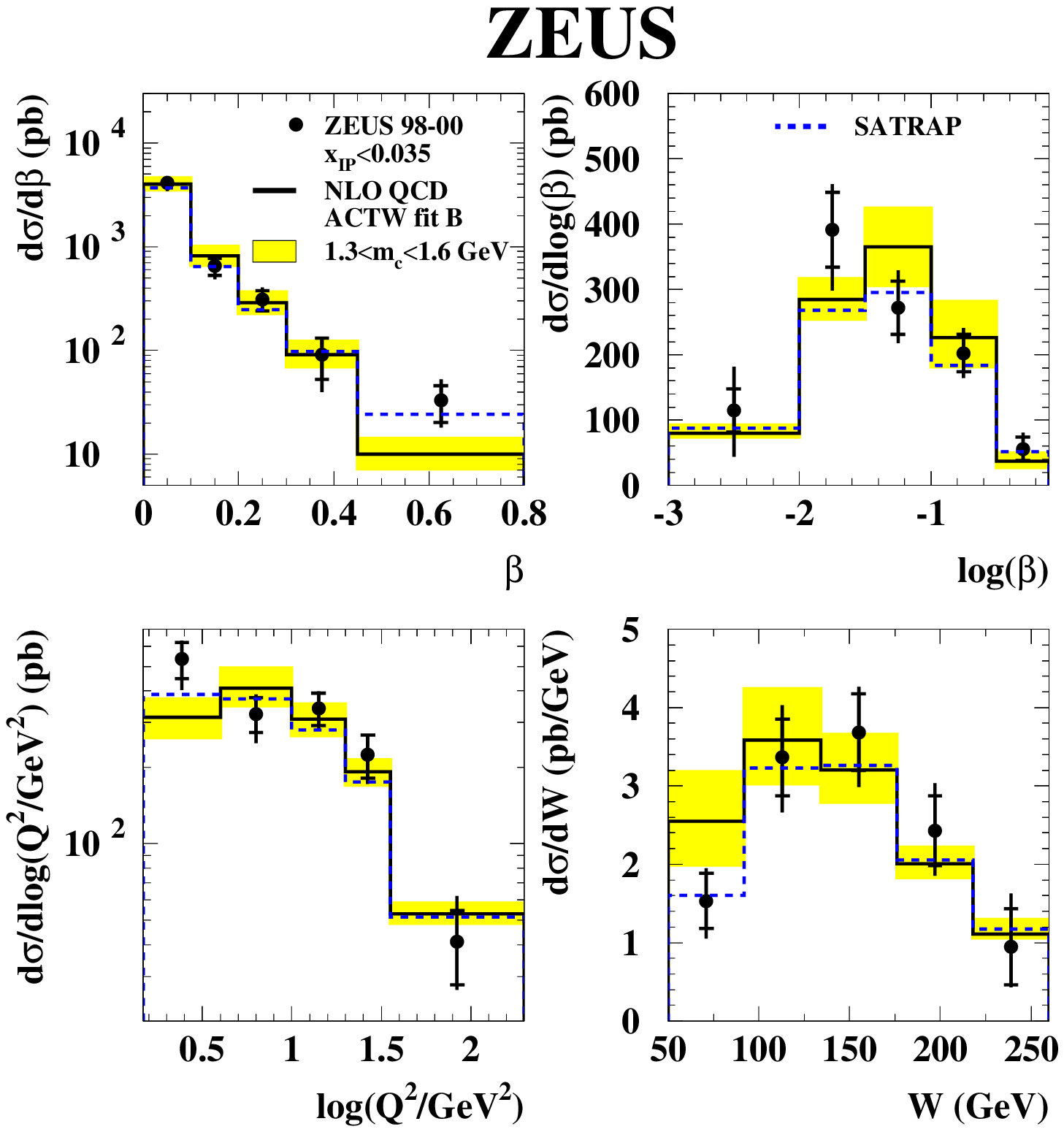}
\end{center}
\caption{Differential cross sections for diffractive $D^{*\pm}$ production
with $x_\pomsub <0.035$ for the data (dots) compared with the
ACTW NLO (solid histogram) and SATRAP (dashed histogram) predictions.
The shaded area shows the effect of
varying the charm-quark mass in the ACTW NLO prediction.
The cross sections are
shown as a function of $\beta$, $\log(\beta)$, $\log(Q^2)$ and $W$.
The inner error bars indicate the
statistical uncertainties, while the outer ones correspond to
statistical and systematic
uncertainties added in quadrature.
The overall normalisation uncertainties arising from
the luminosity measurement ($\pm2.2\%$), from the $\dstar$ and $D^0$ branching
ratios ($\pm2.5\%$) and from the proton-dissociative background
subtraction ($\pm4.9\%$) are not indicated.
}\label{fig-fig6b}
\end{figure}

The differential cross sections as functions of $p_T(D^{*\pm})$,
$\eta(D^{*\pm})$, $\log(M_X^2)$, $x(D^{*\pm})$, $\beta$, $\log(\beta)$,
$\log(Q^2)$ and $W$ were measured for $x_\pomsub < 0.01$ and
$x_\pomsub < 0.035$ (Tables~\ref{tab-xsecsysaxp} and~\ref{tab-xsecsysbxp}).
Figure~\ref{fig-fig7} compares the differential cross sections measured
for $x_\pomsub < 0.01$ with the ACTW, SATRAP and BJLW predictions.
In Figs.~\ref{fig-fig6a} and~\ref{fig-fig6b},
the ACTW and SATRAP predictions are compared with
the differential cross sections measured for $x_\pomsub < 0.035$.

The two-gluon-exchange BJLW model predictions, obtained with
the cutoff value $k_{T,g}^{\rm cut}=1.5\,$GeV tuned using the H1
measurement~\cite{pl:b520:191}, describe the differential cross
sections in the range $x_\pomsub<0.01$ both in shape and normalisation.
Using the value $k_{T,g}^{\rm cut}=1.0\,$GeV ($2.0\,$GeV),
the model predictions significantly overestimate (underestimate)
the data in this range (not shown).

The two-gluon-exchange saturation model (SATRAP) predictions
reproduce the shapes and the normalisations of the differential
cross sections measured in both $x_\pomsub$ ranges.

The ACTW NLO predictions, obtained with the gluon-dominated fit B,
describe the data reasonably well in both $x_\pomsub$ ranges.
Using other gluon-dominated fits, the predictions significantly
overestimate (fit D) or underestimate (fit SG) the data (not shown).
The quark-dominated fits A and C were excluded by the previous ZEUS
measurement~\cite{pl:b545:244}.

\subsection{Ratio of diffractive to inclusive $D^{*\pm}$ production}
\label{sec-rat}

The ratio of diffractively produced $D^{*\pm}$ mesons to inclusive
$D^{*\pm}$ mesons, $R_D$, was measured for $x < 0.028$. This limit
is the product of the $x_\pomsub$ and $\beta$ requirements
imposed for the diffractive $D^{*\pm}$ sample. The ratio of diffractive
to inclusive DIS $D^{*\pm}$ production is then defined by
$$R_D = \frac{\sigma_{ep \rightarrow e D^{*\pm} X' p}(x_\pomsub<0.035,\beta<0.8)}
{\sigma_{ep \rightarrow e D^{*\pm} Y}(x < 0.028)} .$$
Sources of systematic uncertainty in the ratio measurement
were studied in a similar manner to those for the cross-section measurements.
There is a cancellation between the common systematic uncertainties
originating from the selection of inclusive DIS events,
the selection of $\dstar$ candidates and the background estimation.
An additional contribution originates from the model dependence of
the acceptance corrections used in the evaluation of the inclusive
DIS $D^{*\pm}$ cross sections. This systematic uncertainty was estimated
using the inclusive RAPGAP MC sample generated with LEPTO parton showers
instead of the ARIADNE higher-order QCD corrections and with the HERWIG
MC sample.

The ratio measured in the kinematic region $1.5<Q^2<200\gev^2$,
$0.02<y<0.7$, $p_T(D^{*\pm})>1.5\gev$, $|\eta(D^{*\pm})| < 1.5$ and $x<0.028$ is
$$R_D = 6.4\pm0.5(\rm{stat.})^{+0.3}_{-0.7}(\rm{syst.})^{+0.3}_{-0.3}(\rm{p.diss.})~\%.$$
The value is consistent with previous measurements performed in similar
kinematic ranges~\cite{pl:b520:191,pl:b545:244}.

\begin{figure}[tbhp]
\begin{center}
\includegraphics[height=160mm]{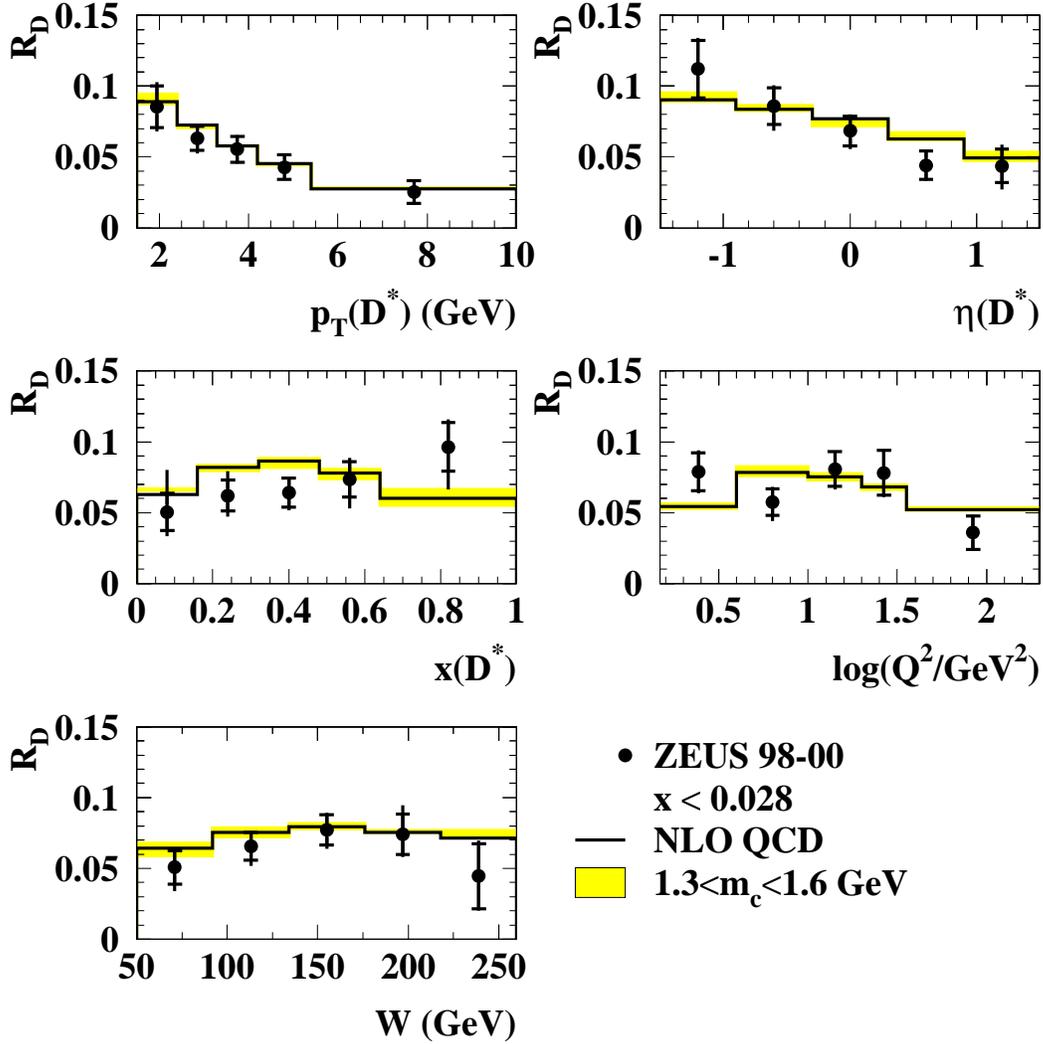}
\end{center}
\caption{The measured ratio of diffractively produced $D^{*\pm}$
mesons to inclusive $D^{*\pm}$ meson production (dots). The ratio
is shown  as a function of $p_T(D^{*\pm})$, $\eta(D^{*\pm})$, $x(D^{*\pm})$,
$\log(Q^2)$ and $W$. The inner error bars indicate the
statistical uncertainties, while the outer ones correspond to
statistical and systematic
uncertainties added in quadrature. The histogram corresponds to the
NLO QCD prediction
where the shaded area shows the effect of
varying the charm-quark mass.
The overall normalisation uncertainty arising from
the proton-dissociative background
subtraction ($\pm4.9\%$) is not indicated.
}\label{fig-fig8}
\end{figure}

Figure~\ref{fig-fig8} (Table~\ref{tab-incsys}) shows the ratio measured
as a function of $p_T(D^{*\pm})$, $\eta(D^{*\pm})$, $x(D^{*\pm})$,
$\log(Q^2)$ and $W$. The measured $R_D$ shows no dependence on $Q^2$,
$W$ or $x(D^{*\pm})$. The relative diffractive contribution is larger
at small $p_T(D^{*\pm})$ and in the backward direction (negative
$\eta(D^{*\pm})$). The NLO QCD predictions for the ratio of diffractive to
inclusive DIS $D^{*\pm}$ production were obtained using ACTW NLO fit B
for the diffractive predictions and the HVQDIS program with the
CTEQ5F3~\cite{epj:c12:375} proton PDF for the inclusive predictions.
Parameters in both calculations were set to the values discussed
in \Sect{theory}. The NLO QCD predictions reproduce the measured $R_D$
values and the trends observed for the $R_D$ distributions measured
as functions of $p_T(D^{*\pm})$ and $\eta(D^{*\pm})$.

\subsection{Open-charm contribution to the diffractive proton structure
              function $F_2^{D(3)}$}
\label{sec-sf}

Neglecting contributions from $Z$-boson exchange and the longitudinal
structure function, the open-charm contribution to the diffractive
structure function of the proton can be related to the cross section,
measured in the full $D^{*\pm}$ kinematic region, by
\begin{equation}
 \frac{1}{2\fcds} \frac{\diff^3\sigma_{ep \rightarrow e D^{*\pm} X' p}}
    {\diff \xpom \diff \beta \diff Q^2} =
    \frac{4\pi \alpha^2_{em}}{Q^4 \beta} (1-y+\frac{y^2}{2})
    F_2^{D(3),c{\bar c}}(\beta,Q^2,\xpom).
\label{eqv-eq1}
\end{equation}

In order to estimate $F_2^{D(3),c{\bar c}}$, the differential cross section
was measured as a function of $\log(\beta)$ for different regions of
$Q^2$ and $\xpom$ (Table~\ref{tab-xsecbt2}). Extrapolation factors of the
measured cross sections to the full $p_T(D^{*\pm})$ and $\eta(D^{*\pm})$
phase space were estimated using the ACTW NLO fit B predictions.
The factors were about five for low-$x_\pomsub$ bins and two for
high-$x_\pomsub$ bins.

In each bin, $F_2^{D(3),c{\bar c}}$ was determined using the formula
$$F_{2~~\rm meas}^{D(3),c{\bar c}}(\beta_i,Q^2_i,x_{\pomsub,i}) =
 \frac{\sigma_{ep \rightarrow e D^{*\pm} X' p}^{i,\rm meas}}
      {\sigma_{ep \rightarrow e D^{*\pm} X' p}^{i,\rm ACTW}}
  F_{2~~\rm ACTW}^{D(3),c{\bar c}}(\beta_i,Q^2_i,x_{\pomsub,i}),$$
where the cross sections $\sigma^i$ in bin $i$ are those
for $p_T(D^{*\pm})>1.5\gev$ and $|\eta(D^{*\pm})| < 1.5$.
The functional form of $F_{2~~\rm ACTW}^{D(3),c{\bar c}}$, calculated 
using Eq.~(\ref{eqv-eq1}), was used
to quote the results for $F_2^{D(3),c{\bar c}}$ at convenient values of
$\beta_i$, $Q^2_i$ and $x_{\pomsub,i}$ close to  the centre-of-gravity of the bin.

The measured $F_2^{D(3),c{\bar c}}$ values are listed in Table~\ref{tab-f2d3}
with their experimental uncertainties. Using ACTW NLO fit D had no
significant effect on the measured values. Other sources of extrapolation
uncertainties are small compared to the experimental
uncertainties~\cite{epj:c12:35}.

\begin{figure}[tbhp]
\begin{center}
\includegraphics[height=160mm]{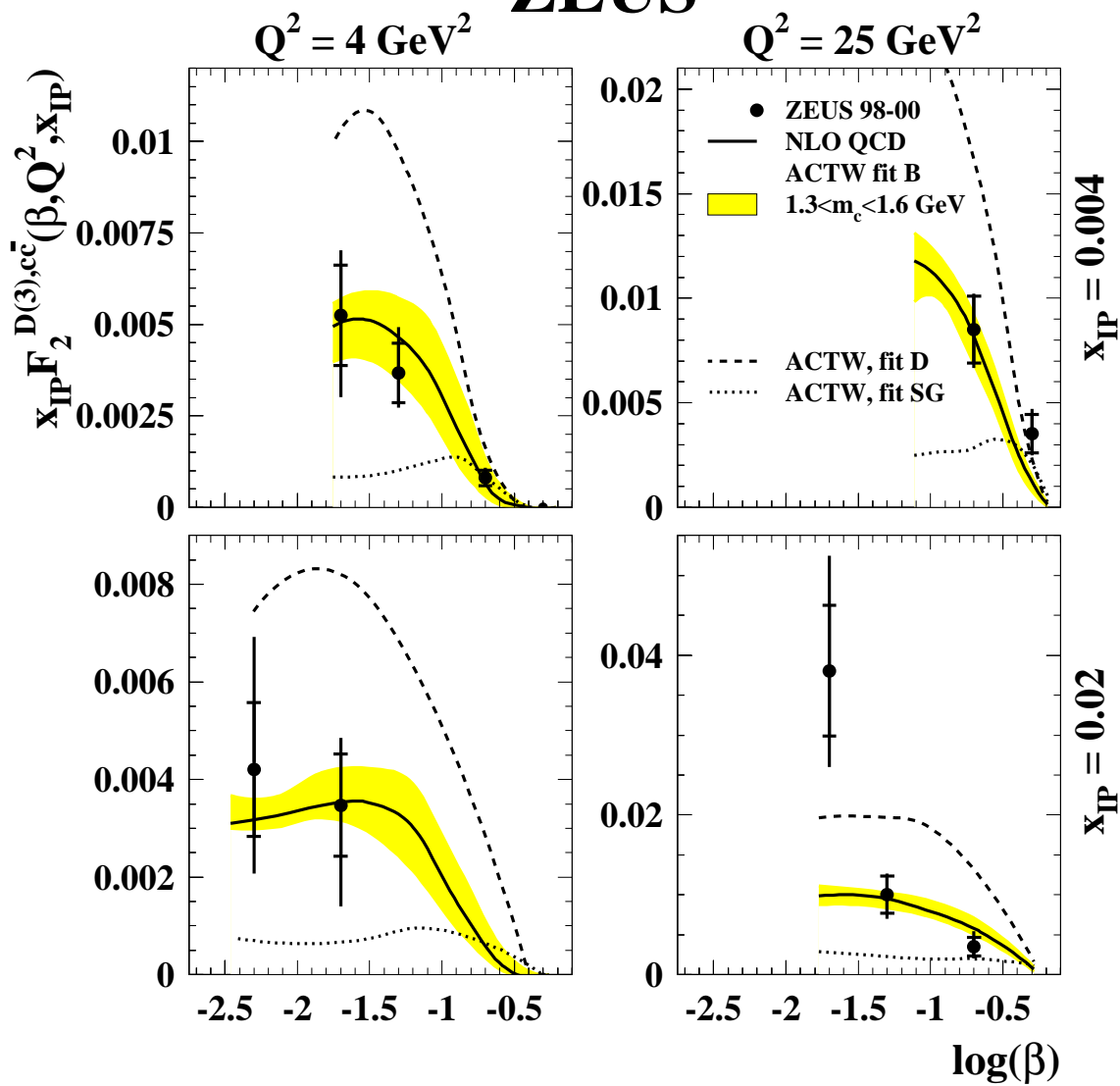}
\end{center}
\caption{The measured charm contribution to the diffractive structure
function of the proton multiplied by $x_\pomsub$,
 $x_\pomsub F_2^{D(3),c{\bar c}}$, as a function of $\beta$
for different values of $Q^2$ and $x_\pomsub$ (dots).
The inner error bars indicate
the statistical uncertainties, while the outer ones correspond to
statistical and systematic
uncertainties added in quadrature.
The overall normalisation uncertainties arising from
the luminosity measurement ($\pm2.2\%$), from the $\dstar$ and $D^0$ branching
ratios ($\pm2.5\%$) and from the proton-dissociative background
subtraction ($\pm4.9\%$) are not indicated.
The curves correspond to the
ACTW model prediction; the shaded area shows the effect of
varying the charm-quark mass.}
\label{fig-fig10}
\end{figure}

Figure~\ref{fig-fig10} shows the quantity $\xpom F_2^{D(3),c{\bar c}}$ as
a function of $\log(\beta)$ for different $Q^2$ and $\xpom$ values.
In all cases, $\xpom F_2^{D(3),c{\bar c}}$ rises as $\beta$ decreases.
The curves show the theoretical $\xpom F_2^{D(3),c{\bar c}}$ obtained using the
ACTW NLO calculations with fit B, D and SG. The fit B prediction generally
agrees with the data. The fit D (SG) prediction overestimates (underestimates)
the measured $\xpom F_2^{D(3),c{\bar c}}$ at low $\beta$.

\section{Summary}
\label{sec-sum}

Diffractive $D^{*\pm}$ production has been measured in the kinematic region
$1.5 < Q^2 < 200\gev^2$, $0.02 < y < 0.7$, $x_\pomsub < 0.035$, $\beta < 0.8$,
$p_T(D^{*\pm}) >1.5\gev$ and $|\eta(D^{*\pm})| < 1.5$.
The cross section integrated over this kinematic region is
$521 \pm 43(\rm{stat.})^{+34}_{-58}(\rm{syst.}) \pm 26(\rm{p.diss.})$~pb.
Differential cross sections have been compared to
the predictions of different diffractive models.
The ACTW NLO predictions, based on parton densities of the Pomeron
obtained from combined fits to the inclusive diffractive
DIS and diffractive dijet photoproduction measurements at HERA,
describe the results reasonably well in the whole $x_\pomsub$ range if
the gluon-dominated fit B is used.
The predictions of the two-gluon-exchange saturation model also
reproduce the shapes and normalisations of the differential
cross sections in the whole $x_\pomsub$ range.
The predictions of the two-gluon-exchange BJLW model describe
the cross sections measured for $x_\pomsub<0.01$,
if a minimum value for the transverse momentum of the final-state gluon of
$k_{T,g}^{\rm cut}=1.5\,$GeV is used.

The ratio of diffractive $\,\,\,D^{*\pm}$ $\,\,\,$production $\,\,\,$to $\,\,\,$inclusive
$\,\,\,$DIS $\,\,\,$$D^{*\pm}$ $\,\,\,$production $\,\,\,$
has been measured to be
$R_D = 6.4 \pm 0.5(\rm{stat.})^{+0.3}_{-0.7}(\rm{syst.})^{+0.3}_{-0.3}(\rm{p.diss.})~\%$.
The ratio $R_D$ shows no dependence on $W$, $Q^2$ or $x(D^{*\pm})$.
The relative contribution from diffraction is larger at small $p_T(D^{*\pm})$
and in the backward direction (negative $\eta(D^{*\pm})$).
The NLO QCD predictions reproduce the measured $R_D$.

The open-charm contribution, $F_2^{D(3),c{\bar c}}$, to the diffractive proton
structure function has been extracted. For all values of $Q^2$ and $\xpom$,
$F_2^{D(3),c{\bar c}}$ rises as $\beta$ decreases.
The results have been compared with the
theoretical $F_2^{D(3),c{\bar c}}$ obtained using the
ACTW NLO calculations with the gluon-dominated fits B, D and SG.
The data exclude the fits D and SG, and are consistent with fit B.
This demonstrates that the data have a strong sensitivity to the diffractive parton
densities, and that diffractive PDFs in NLO QCD are able to consistently
describe both inclusive diffractive DIS and diffractive charm production in DIS.

\section{Acknowledgments}
\label{sec-sum}

We would like to thank the DESY Directorate for their strong support
and encouragement. The remarkable achievements of the HERA machine
group were vital for the successful completion of this work
and are greatly appreciated. The design, construction and installation
of the ZEUS detector have been made possible by the ingenuity and 
effort of many people who are not listed as authors. 
We thank J.~Bartels and H.~Jung for informative discussions. 

\vfill\eject

{
\def\bibname{\Large\bf References}
\def\refname{\Large\bf References}
\pagestyle{plain}
\ifzeusbst
  \bibliographystyle{./BiBTeX/bst/l4z_default}
\fi
\ifzdrftbst
  \bibliographystyle{./BiBTeX/bst/l4z_draft}
\fi
\ifzbstepj
  \bibliographystyle{./BiBTeX/bst/l4z_epj}
\fi
\ifzbstnp
  \bibliographystyle{./BiBTeX/bst/l4z_np}
\fi
\ifzbstpl
  \bibliographystyle{./BiBTeX/bst/l4z_pl}
\fi
{\raggedright
\bibliography{./BiBTeX/user/syn.bib,%
              ./BiBTeX/bib/l4z_articles.bib,%
              ./BiBTeX/bib/l4z_books.bib,%
              ./BiBTeX/bib/l4z_conferences.bib,%
              ./BiBTeX/bib/l4z_h1.bib,%
              ./BiBTeX/bib/l4z_misc.bib,%
              ./BiBTeX/bib/l4z_old.bib,%
              ./BiBTeX/bib/l4z_preprints.bib,%
              ./BiBTeX/bib/l4z_replaced.bib,%
              ./BiBTeX/bib/l4z_temporary.bib,%
              ./BiBTeX/bib/l4z_zeus.bib}}
}
\vfill\eject

%
\begin{table}
\begin{center}
\begin{tabular}{|l|c|} \hline
$\xpom$ bin & $\diff \sigma / \diff \xpom$ (nb) \\ \hline
$0$    ,  $0.003$    &$28.0\pm4.9_{-3.2}^{+3.4}$  \\ \hline 
$0.003$ , $0.006$    &$25.4\pm4.7_{-2.4}^{+5.1}$  \\ \hline 
$0.006$ , $0.010$    &$18.6\pm3.6_{-2.5}^{+1.9}$  \\ \hline 
$0.010$ , $0.020$    &$13.7\pm2.2_{-2.3}^{+1.6}$  \\ \hline 
$0.020$ , $0.035$    &$13.7\pm2.4_{-2.9}^{+5.0}$  \\ \hline 
\end{tabular}                              
\caption{Differential cross section for diffractive $D^{*\pm}$ production
as a function of $\xpom$.
The first and second uncertainties represent
statistical and systematic uncertainties, respectively.
The overall normalisation uncertainties arising from
the luminosity measurement ($\pm2.2\%$), from the $\dstar$ and $D^0$ branching
ratios ($\pm2.5\%$) and from the proton-dissociative background
subtraction ($\pm4.9\%$) are not indicated.
}
\label{tab-xsecxp}
\end{center}
\end{table}
\begin{table}
\begin{center}
\begin{tabular}{|l|c|c|} \hline
$p_T(\dstar)$ bin ($\gev$)  & \multicolumn{2}{|c|}{$\diff \sigma / \diff p_T (\dstar)$ (pb$/\gev$)}  \\ \hline
& $\xpom < 0.01$ & $\xpom < 0.035$ \\ \hline
$1.5$ , $2.4$      &$161\pm29_{-19}^{+28}$ &$307\pm50 _{-42 }^{+44 }$ \\ \hline
$2.4$ , $3.3$      &$66 \pm11_{-7 }^{+8 }$ &$151\pm20 _{-19 }^{+16 }$ \\ \hline
$3.3$ , $4.2$      &$19 \pm 5_{-2 }^{+2 }$ &$70 \pm11 _{-7  }^{+4  }$ \\ \hline 
$4.2$ , $5.4$      &$10 \pm 3_{-1 }^{+1 }$ &$26 \pm5  _{-2  }^{+3  }$ \\ \hline 
$5.4$ , $10.0$     &                       &$2.8\pm0.9_{-0.5}^{+0.3}$ \\ \hline 
\hline
$\eta(\dstar)$ bin & \multicolumn{2}{|c|}{$\diff \sigma / \diff \eta (\dstar)$ (pb)} \\ \hline
& $\xpom < 0.01$ & $\xpom < 0.035$ \\ \hline
$-1.5$ , $-0.9$   &$124\pm26_{-16}^{+13}$ &$212\pm36_{-27}^{+27}$    \\ \hline 
$-0.9$ , $-0.3$   &$104\pm19_{-6 }^{+14}$ &$213\pm31_{-30}^{+28}$    \\ \hline 
$-0.3$ , $0.3$    &$78 \pm17_{-9 }^{+11}$ &$195\pm29_{-27}^{+32}$    \\ \hline 
$0.3$ , $0.9$     &$37 \pm13_{-12}^{+8 }$ &$125\pm28_{-29}^{+18}$    \\ \hline 
$0.9$ , $1.5$     &$55 \pm20_{-11}^{+21}$ &$134\pm36_{-38}^{+38}$    \\ \hline 
\hline
$\log(M_X^2/\gev^2)$ bin & \multicolumn{2}{|c|}{$\diff \sigma / \diff \log(M_X^2/\gev^2)$ (pb)} \\ \hline
& $\xpom < 0.01$ & $\xpom < 0.035$ \\ \hline
$1.00$ , $1.44$     &$89 \pm21_{-17}^{+21}$ &$94 \pm23_{-21 }^{+22 }$    \\ \hline
$1.44$ , $1.88$     &$195\pm35_{-25}^{+26}$ &$201\pm38_{-28 }^{+22 }$    \\ \hline
$1.88$ , $2.32$     &$200\pm29_{-21}^{+24}$ &$382\pm45_{-46 }^{+37 }$    \\ \hline
$2.32$ , $2.76$     &$47 \pm25_{-16}^{+17}$ &$284\pm54_{-60 }^{+41 }$    \\ \hline
$2.76$ , $3.20$     &                       &$286\pm65_{-102}^{+129}$    \\ \hline
\hline
$x(D^{*\pm})$ bin & \multicolumn{2}{|c|}{$\diff \sigma / \diff x(D^{*\pm})$  (pb)} \\ \hline
& $\xpom < 0.01$ & $\xpom < 0.035$ \\ \hline
$0.00$ , $0.16$      &$185\pm61_{-43}^{+62}$ &$429\pm107_{-125}^{+161}$  \\ \hline 
$0.16$ , $0.32$      &$252\pm76_{-52}^{+74}$ &$788\pm135_{-156}^{+163}$  \\ \hline 
$0.32$ , $0.48$      &$446\pm85_{-46}^{+39}$ &$864\pm134_{-121}^{+76 }$  \\ \hline 
$0.48$ , $0.64$      &$376\pm75_{-78}^{+67}$ &$726\pm119_{-157}^{+106}$  \\ \hline 
$0.64$ , $1.00$      &$92 \pm21_{-9 }^{+18}$ &$221\pm38 _{-39 }^{+27 }$  \\ \hline 
\end{tabular}                              
\caption{Differential cross sections for diffractive $D^{*\pm}$ production
as a function of $p_T(\dstar)$, $\eta(\dstar)$, $\log(M_X^2)$ and
$x(D^{*\pm})$.
The first and second uncertainties are
statistical and systematic, respectively.
The overall normalisation uncertainties arising from
the luminosity measurement ($\pm2.2\%$), from the $\dstar$ and $D^0$ branching
ratios ($\pm2.5\%$) and from the proton-dissociative background
subtraction ($\pm4.9\%$) are not indicated.
}
\label{tab-xsecsysaxp}
\end{center}
\end{table}
\begin{table}
\begin{center}
\begin{tabular}{|l|c|c|} \hline
$\beta$ bin & \multicolumn{2}{|c|}{$\diff \sigma / \diff \beta$ (pb)} \\ \hline
& $\xpom < 0.01$ & $\xpom < 0.035$ \\ \hline
$0.00$ , $0.10$          &$1252\pm203_{-118}^{+170}$ &$4153\pm410_{-558}^{+243}$   \\ \hline 
$0.10$ , $0.20$          &$419 \pm94 _{-56 }^{+32 }$ &$654\pm125 _{-113}^{+125}$   \\ \hline 
$0.20$ , $0.30$          &$244 \pm54 _{-20 }^{+40 }$ &$311\pm69  _{-38 }^{+62 }$   \\ \hline 
$0.30$ , $0.45$          &$100 \pm35 _{-27 }^{+15 }$ &$91\pm39   _{-34 }^{+22 }$   \\ \hline 
$0.45$ , $0.80$          &$27  \pm11 _{-5  }^{+14 }$ &$33\pm13   _{-8  }^{+15 }$   \\ \hline 
\hline
$\log(\beta)$ bin & \multicolumn{2}{|c|}{$\diff \sigma / \diff \log(\beta)$ (nb)} \\ \hline
& $\xpom < 0.01$ & $\xpom < 0.035$ \\ \hline
$-3.0$ , $-2.0$    &                       &$115\pm33_{-63}^{+58}$ \\ \hline 
$-2.0$ , $-1.5$    &$105\pm28_{-33}^{+22}$ &$392\pm58_{-74}^{+39}$ \\ \hline 
$-1.5$ , $-1.0$    &$124\pm25_{-17}^{+27}$ &$272\pm41_{-37}^{+40}$ \\ \hline 
$-1.0$ , $-0.5$    &$141\pm22_{-13}^{+12}$ &$203\pm28_{-26}^{+26}$ \\ \hline 
$-0.5$ , $-0.1$    &$65 \pm16_{-11}^{+14}$ &$56 \pm18_{-9 }^{+17}$ \\ \hline 
\hline
$\log(Q^2/\gev^2)$ bin & \multicolumn{2}{|c|}{$\diff \sigma / \diff \log(Q^2/\gev^2)$ (pb)} \\ \hline
& $\xpom < 0.01$ & $\xpom < 0.035$ \\ \hline
$0.17$, $0.6$   &$276\pm51_{-34}^{+51}$ &$534\pm87_{-96}^{+46}$   \\ \hline 
$0.6$ , $1.0$     &$140\pm29_{-15}^{+26}$ &$324\pm51_{-55}^{+35}$   \\ \hline 
$1.0$ , $1.3$     &$106\pm27_{-6 }^{+8 }$ &$342\pm50_{-34}^{+28}$   \\ \hline 
$1.3$ , $1.55$   &$103\pm25_{-10}^{+10}$ &$225\pm43_{-29}^{+13}$   \\ \hline 
$1.55$ , $2.3$   &$17 \pm7 _{-3 }^{+4 }$ &$41\pm13_ {-6 }^{+16}$   \\ \hline 
\hline
$W$ bin ($\gev$) & \multicolumn{2}{|c|}{$\diff \sigma / \diff W$ (pb/\gev)} \\ \hline
& $\xpom < 0.01$ & $\xpom < 0.035$ \\ \hline
$50$ ,  $92$    &$0.45\pm0.14_{-0.09}^{+0.13}$ &$1.53\pm0.35_{-0.33}^{+0.23}$   \\ \hline 
$92$ ,  $134$   &$1.48\pm0.29_{-0.21}^{+0.23}$ &$3.36\pm0.49_{-0.51}^{+0.45}$   \\ \hline 
$134$ , $176$   &$1.63\pm0.29_{-0.21}^{+0.16}$ &$3.68\pm0.49_{-0.50}^{+0.32}$   \\ \hline 
$176$ , $218$   &$1.25\pm0.29_{-0.12}^{+0.25}$ &$2.43\pm0.44_{-0.37}^{+0.41}$   \\ \hline 
$218$ , $260$   &$0.50\pm0.33_{-0.15}^{+0.22}$ &$0.95\pm0.48_{-0.18}^{+0.48}$   \\ \hline
\end{tabular}                              
\caption{Differential cross sections for diffractive $D^{*\pm}$ production
as a function of $\beta$, $\log(\beta)$, $\log(Q^2)$ and $W$.
The first and second uncertainties are
statistical and systematic, respectively.
The overall normalisation uncertainties arising from
the luminosity measurement ($\pm2.2\%$), from the $\dstar$ and $D^0$ branching
ratios ($\pm2.5\%$) and from the proton-dissociative background
subtraction ($\pm4.9\%$) are not indicated.
}
\label{tab-xsecsysbxp}
\end{center}
\end{table}
\begin{table}
\begin{center}
\begin{tabular}{|l|c|} \hline
$p_T(\dstar)$ bin ($\gev$)  & $R_D$ ($\%$)  \\ \hline
$1.5$ , $2.4$      &$8.5\pm1.5_{-0.9}^{+0.9}$    \\ \hline 
$2.4$ , $3.3$      &$6.3\pm0.9_{-0.7}^{+0.2}$    \\ \hline 
$3.3$ , $4.2$      &$5.5\pm0.9_{-0.5}^{+0.3}$    \\ \hline 
$4.2$ , $5.4$      &$4.3\pm0.9_{-0.2}^{+0.4}$    \\ \hline 
$5.4$ , $10.0$     &$2.5\pm0.8_{-0.4}^{+0.2}$    \\ \hline 
\hline
$\eta(\dstar)$ bin & $R_D$ ($\%$) \\ \hline
$-1.5$ , $-0.9$   &$11.2\pm2.0_{-0.9}^{+0.9}$   \\ \hline 
$-0.9$ , $-0.3$   &$8.6\pm1.3_{-1.1}^{+0.7}$   \\ \hline
$-0.3$ , $0.3$    &$6.8\pm1.1_{-0.7}^{+0.5}$   \\ \hline
$0.3$ , $0.9$     &$4.4\pm1.0_{-0.4}^{+0.5}$   \\ \hline
$0.9$ , $1.5$     &$4.4\pm1.2_{-1.1}^{+0.9}$   \\ \hline
\hline
$x(D^{*\pm})$ bin & $R_D$ ($\%$) \\ \hline
$0.00$ , $0.16$      &$5.0\pm1.3_{-1.1}^{+2.7}$    \\ \hline 
$0.16$ , $0.32$      &$6.2\pm1.1_{-1.0}^{+1.3}$    \\ \hline 
$0.32$ , $0.48$      &$6.4\pm1.0_{-0.7}^{+0.4}$    \\ \hline 
$0.48$ , $0.64$      &$7.4\pm1.2_{-1.6}^{+0.9}$    \\ \hline 
$0.64$ , $1.00$      &$9.6\pm1.7_{-2.5}^{+0.9}$    \\ \hline 
\hline
$\log(Q^2/\gev^2)$ bin & $R_D$ ($\%$) \\ \hline
$0.17$ , $0.60$    &$7.9\pm1.3_{-0.7}^{+0.7}$    \\ \hline 
$0.60$ , $1.00$    &$5.8\pm0.9_{-1.0}^{+0.5}$    \\ \hline 
$1.00$ , $1.30$    &$8.1\pm1.2_{-0.7}^{+0.4}$    \\ \hline 
$1.30$ , $1.55$    &$7.8\pm1.6_{-0.7}^{+0.2}$    \\ \hline 
$1.55$ , $2.30$    &$3.6\pm1.2_{-0.4}^{+0.4}$    \\ \hline 
\hline
$W$ bin ($\gev$) & $R_D$ ($\%$) \\ \hline
$50$ , $92$      &$5.1\pm1.2_{-1.2}^{+0.3}$    \\ \hline 
$92$ , $134$     &$6.6\pm1.0_{-1.0}^{+0.4}$    \\ \hline 
$134$ , $176$    &$7.7\pm1.1_{-0.8}^{+0.6}$    \\ \hline 
$176$ , $218$    &$7.4\pm1.4_{-0.7}^{+1.4}$    \\ \hline 
$218$ , $260$    &$4.4\pm2.3_{-0.8}^{+1.0}$    \\ \hline
\end{tabular}                              
\caption{Ratio of diffractively produced $D^{*\pm}$
mesons to inclusive $D^{*\pm}$ meson production
as a function of $p_T(D^{*\pm})$, $\eta(D^{*\pm})$, $x(D^{*\pm})$,
$\log(Q^2)$ and $W$.
The first and second uncertainties are
statistical and systematic, respectively.
The overall normalisation uncertainties arising from
the luminosity measurement ($\pm2.2\%$), from the $\dstar$ and $D^0$ branching
ratios ($\pm2.5\%$) and from the proton-dissociative background
subtraction ($\pm4.9\%$) are not indicated.
}
\label{tab-incsys}
\end{center}
\end{table}
\begin{table}
\begin{center}
\begin{tabular}{|l|c|c|} \hline
$\log(\beta)$ bin    & \multicolumn{2}{|c|}{$\diff \sigma / \diff \log(\beta)$, $\xpom < 0.01$ (pb)} \\ \hline
& $1.5 < Q^2 < 10$ $\gev^2$       & $10 < Q^2 < 200$ $\gev^2$     \\ \hline
$-2.0$ , $-1.5$   &$107\pm28_{-36}^{+23}$  &                       \\ \hline
$-1.5$ , $-1.0$   &$114\pm25_{-16}^{+30}$  &                       \\ \hline
$-1.0$ , $-0.5$   &$62 \pm16_{-8 }^{+14}$  &$80\pm15_{-9}^{+6 }$  \\ \hline
$-0.5$ , $-0.1$   &                        &$61\pm16_{-9}^{+13}$  \\ \hline
\hline
$\log(\beta)$ bin    & \multicolumn{2}{|c|}{$\diff \sigma / \diff \log(\beta)$, $0.01<\xpom < 0.035$ (pb)}  \\ \hline
& $1.5 < Q^2 < 10$ $\gev^2$       & $10 < Q^2 < 200$ $\gev^2$     \\ \hline
$-3.0$ , $-2.0$   &$96 \pm31_{-37}^{+53}$  &                        \\ \hline
$-2.0$ , $-1.5$   &$142\pm43_{-73}^{+36}$  &$141\pm30_{-32}^{+44}$  \\ \hline
$-1.5$ , $-1.0$   &                        &$106\pm25_{-20}^{+13}$  \\ \hline
$-1.0$ , $-0.5$   &                        &$52 \pm17_{-14}^{+22}$  \\ \hline
\end{tabular}                                                              
\caption{Differential cross section for diffractive $D^{*\pm}$ production
as a function of $\log(\beta)$
for different regions of $Q^2$ and $\xpom$.
The first and second uncertainties are
statistical and systematic, respectively.
The overall normalisation uncertainties arising from
the luminosity measurement ($\pm2.2\%$), from the $\dstar$ and $D^0$ branching
ratios ($\pm2.5\%$) and from the proton-dissociative background
subtraction ($\pm4.9\%$) are not indicated.
}
\label{tab-xsecbt2}
\end{center}
\end{table}
\begin{table}
\begin{center}
\begin{tabular}{|l|c|c|} \hline
\multicolumn{3}{|c|}{$F_2^{D(3)c\bar{c}}$, $\xpom = 0.004$} \\ \hline
$\beta$    &$Q^2 = 4\gev^2$       &$Q^2 = 25\gev^2$     \\ \hline
$0.020$   &$1.34\pm0.35_{-0.44}^{+0.28}$  &                                \\ \hline
$0.050$   &$0.92\pm0.20_{-0.13}^{+0.24}$  &                                \\ \hline
$0.200$   &$0.20\pm0.05_{-0.03}^{+0.05}$  &$2.14\pm0.40_{-0.23}^{+0.16}$  \\ \hline
$0.500$   &                               &$0.89\pm0.23_{-0.12}^{+0.18}$  \\ \hline
\hline
\multicolumn{3}{|c|}{$F_2^{D(3)c\bar{c}}$, $\xpom = 0.02$}  \\ \hline
$\beta$    &$Q^2 = 4\gev^2$           &$Q^2 = 25\gev^2$              \\ \hline
$0.005$   &$0.20\pm0.07_{-0.08}^{+0.12}$  &                                  \\ \hline
$0.020$   &$0.17\pm0.05_{-0.09}^{+0.04}$  &$1.87\pm0.40_{-0.44}^{+0.59}$  \\ \hline
$0.050$   &                               &$0.50\pm0.12_{-0.09}^{+0.06}$  \\ \hline
$0.200$   &                               &$0.18\pm0.06_{-0.05}^{+0.08}$  \\ \hline
\end{tabular}                                                              
\caption{The measured charm contribution to the diffractive structure
function of the proton,
$F_2^{D(3),c{\bar c}}$,
for different values of $\beta$, $Q^2$ and $x_\pomsub$.
The first and second uncertainties are
statistical and systematic, respectively.
The overall normalisation uncertainties arising from
the luminosity measurement ($\pm2.2\%$), from the $\dstar$ and $D^0$ branching
ratios ($\pm2.5\%$) and from the proton-dissociative background
subtraction ($\pm4.9\%$) are not indicated.
}
\label{tab-f2d3}
\end{center}
\end{table}


\newpage
\newpage
\newpage
\newpage
\newpage
\newpage
\newpage
\newpage
\newpage
\newpage
\newpage
%

%
%
\end{document}